%% file: Sommer2023.tex
\documentclass[switch]{article} 
\usepackage{preprint}

\usepackage{graphicx}      
\usepackage{natbib}        
\usepackage{amssymb}	
\usepackage{amsmath}
\usepackage[ruled,vlined,algo2e]{algorithm2e} 
\usepackage{xcolor}
\usepackage{import}
\usepackage{mathrsfs}
\usepackage{bm}
\usepackage{multirow}
\usepackage{placeins}
\usepackage{afterpage}

\usepackage[font=footnotesize]{caption}

\usepackage[utf8]{inputenc}	
\usepackage[T1]{fontenc}	
\usepackage{xcolor}		
\usepackage[colorlinks = true,
            linkcolor = purple,
            urlcolor  = blue,
            citecolor = cyan,
            anchorcolor = black]{hyperref}	
\usepackage{booktabs} 		
\usepackage{nicefrac}		
\usepackage{microtype}		
\usepackage{lineno}		
\usepackage{float}			

\usepackage{tikz,xcolor,hyperref}
\definecolor{lime}{HTML}{A6CE39}
\DeclareRobustCommand{\orcidicon}{
	\begin{tikzpicture}
	\draw[lime, fill=lime] (0,0) 
	circle [radius=0.16] 
	node[white] {{\fontfamily{qag}\selectfont \tiny ID}};
	\draw[white, fill=white] (-0.0625,0.095) 
	circle [radius=0.007];
	\end{tikzpicture}
	\hspace{-2mm}
}
\foreach \x in {A, ..., Z}{\expandafter\xdef\csname orcid\x\endcsname{\noexpand\href{https://orcid.org/\csname orcidauthor\x\endcsname}
			{\noexpand\orcidicon}}
}

\usepackage{tabularx} 
\usepackage{mdframed}
\usepackage{multicol}
\usepackage{nomencl}
\usepackage{etoolbox}
\renewcommand\nomgroup[1]{%
  \item[\bfseries
  \ifstrequal{#1}{A}{Greek letters}{%
  \ifstrequal{#1}{B}{Abbreviations}{}}%
]}
\renewcommand{\nompreamble}{\begin{multicols}{2}}
\renewcommand{\nompostamble}{\end{multicols}}
\makenomenclature

\graphicspath{{figures/}}

\DeclareMathOperator*{\argminA}{arg\,min}
\DeclareMathOperator*{\maxA}{max}
\DeclareMathOperator*{\minA}{min}

\newcommand{\CommentMmo}[1]{\textcolor{red}{ #1 }}
\renewcommand{\CommentMmo}[1]{}

\newcommand{\CommentMark}[1]{\textcolor{red}{#1}}
\renewcommand{\CommentMark}[1]{{#1}}

\title{Estimating flow fields with Reduced Order Models}

\usepackage{authblk}

\author[1\thanks{\tt{Kamil.Sommer@rub.de}}]{Kamil David Sommer}
\author[1]{Lucas Reineking}
\author[2]{Yogesh Parry Ravichandran}
\author[2]{Romuald Skoda}
\author[1]{Martin M\"onnigmann}

\affil[1]{Automatic Control and Systems Theory, Ruhr-Universit\"at Bochum, Bochum, Germany}
\affil[2]{Hydraulic Fluid Machinery, Ruhr-Universit\"at Bochum, Bochum, Germany}

\begin{document}

\maketitle

\begin{abstract}
The estimation of fluid flows inside a centrifugal pump in realtime is a challenging task that cannot be achieved with long-established methods like CFD due to their computational demands. We use a projection-based reduced order model (ROM) instead. Based on this ROM, a realtime observer can be devised that estimates the temporally and spatially resolved velocity and pressure fields inside the pump. The entire fluid-solid domain is treated as a fluid in order to be able to consider moving rigid bodies in the reduction method. A greedy algorithm is introduced for finding suitable and as few measurement locations as possible. Robust observability is ensured with an extended Kalman filter, which is based on a time-variant observability matrix obtained from the nonlinear velocity ROM. We present the results of the velocity and pressure ROMs based on a unsteady Reynolds-averaged Navier-Stokes CFD simulation of a 2D centrifugal pump, as well as the results for the extended Kalman filter.
\end{abstract}
\vspace{0.35cm}


\input{nomenclature.tex}

\section{Introduction}\label{sec:Introduction}
Monitoring the state, i.e., the spatial and temporal velocity and pressure fields, of hydraulic machines such as centrifugal pumps in realtime is a very demanding task~(see, e.g., \citep{Hayase2015}). Reduced order models (ROMs) can provide the same spatial and temporal resolution as computational fluid dynamics (CFD) simulations at a fraction of their computational effort. 
Consequently, ROMs are an ideal basis for methods for the reconstruction of fluid flow and pressure fields in realtime. 

\CommentMark{Reduced order models have been designed for centrifugal pumps before.
The authors in \citep{Wei2023} conducted several stationary CFD-simulations for various operating points, specifically for different rotor rotation speeds and flow rates. Subsequently, they derived a proper orthogonal decomposition (POD) reduced order model, which is designed to reflect the number of distinct operating point variants. In contrast to the present article, this POD-ROM model was used to predict stationary flow fields for various operating points by linear interpolation of the modal coefficients. 
In \citep{dAgostino2012} and \citep{dAgostino2011}, the authors established a reduced order model consisting of partial differential and algebraic equations tailored to centrifugal pumps. This ROM was developed under simplifying assumptions including irrotational flow and inviscidity of the fluid. Each component of the pump was analyzed independently, and distinct models were formulated for the fluid flow within each component. While the model successfully incorporates hydraulic losses, it does so by employing empirical correlations derived from experimental or numerical data. All of these methodologies are limited to stationary flow fields. In contrast, our primary focus lies in the examination and analysis of the unsteady, time-varying flow fields.}

\CommentMark{Reduced order models that combine proper orthogonal decomposition and Galerkin projection (GP) can be used to generate dynamic models capable of computing not only steady-state, but also unsteady flow fields. To the best of the authors' knowledge, there is currently no available literature addressing the utilization of proper orthogonal decomposition and Galerkin projection based reduced order models (POD-GP-ROMs) specifically applied to real-world centrifugal pumps. However, POD-GP-ROMs} have successfully been applied to numerous other problems, e.g., to oscillating and circular cylinders and grooved channels~\citep{Liberge2010, Deane1991,  Bergmann2008}, to magneto-mechanical problems for magnetic resonance imaging~\citep{Seoane2020}, to the flow inside of positive replacement pumps~\citep{Gunder2018}, transient thermal flows in integrated circuits~\citep{Meyer2017}, and to diffusion and heat conduction problems in drying processes~\citep{Berner2017}. ROMs for pressure fields can be constructed with similar methods as for velocity fields~(see, e.g., \citep{Noack2005, Caiazzo2014, Akhtar2009}).

Once a ROM is available, it remains to answer the question how to reconstruct the flow and pressure fields of the actual system that has been modeled. 
System theoretic notions, such as observability and reconstructability, can be used to verify whether a set of local measurements allows to determine the entire velocity and pressure fields. If such a set of measurements has been identified, Kalman filters or their extensions to nonlinear systems can be applied to reconstruct the desired fields in realtime. 
State estimation with reduced order models has successfully been implemented for various problems like contaminant flows~\citep{John2010}, cavity flow oscillations~\citep{Rowley2005}, positive displacement pumps~\citep{Gunder2018}, and reaction-diffusion processes~\citep{Berner2020}. 

\CommentMark{It is the main contribution of this paper to combine classical POD-Galerkin reduced order models and an extended Kalman filter. A greedy algorithm is used to identify optimal measurement locations. Our approach is not limited to be used for centrifugal pumps only but can be extended to address a wide range of fluid flow problems.} We show that the complete pressure and velocity fields can be monitored with velocity measurements at only a few measurement locations and an extended Kalman filter. Moreover, we show that a stable estimation is possible based on ROMs that require a much lower computational effort than the original CFD simulation, which is a crucial step towards practically relevant methods for realtime monitoring of velocity and pressure fields in pumps.
We \CommentMark{use classical projection based ROM methods ~(see, e.g., \citep{Deane1991,John2010}) to derive reduced order models} for a 2D intersection of a realistic centrifugal pump and are mainly interested in the incompressible velocity and pressure field. We generate a set of snapshot data based on unsteady Reynolds-averaged Navier-Stokes (URANS) CFD solutions, based on which we construct the ROMs. A projection-based model reduction transforms the underlying partial differential equations (Navier-Stokes and Pressure-Poisson equations) into a set of ordinary differential and algebraic equations. For the sake of simplicity, we use a finite difference discretization scheme instead of finite elements or finite volumes for the reduced order model (see, e.g., \citep{Lorenzi2016}). Fluid-structure interaction in centrifugal pumps poses a difficulty for our reduction method. Systems with moving or deforming grids have  been investigated before~\citep{Liberge2010, Falaize2019, Xu2020, Ballarin2016, Placzek2011}. In most of these works, either the 3D-CFD simulation itself was carried out in a fixed stationary grid using methods such as the immersed boundary method or the fictitious domain method~\citep{Court2014, Fadlun2000}, or the solution obtained on a moving grid was transferred in a post-processing step to the fixed stationary grid. These approaches often require a special treatment of moving and deforming solid domains in the reduced order modeling method. We use a simple but effective approach and treat the complex-body motion as a fluid domain with artificial flow fields by interpolating the values between suction and pressure side of the impeller blades \CommentMark{using the smoothing and interpolation method from~\citep{Garcia2010}.} As a result, the model order reduction may be carried out on a fixed stationary grid and well-established POD and Galerkin projection methods for fixed boundaries can be applied. 

\CommentMark{The employed observer relies on two key ingredients: the reduced order model and the extended Kalman filter.} In comparison to CFD models, the resulting ROMs can be solved with significantly less computational effort \CommentMark{and thus serve as the dynamic model}. This allows us to apply an extended Kalman filter to estimate the state, i.e., to determine the velocity and pressure fields, based on flow vectors at a few measurement locations. \CommentMark{An accurate but not necessarily stable reduced order model is required to use the extended Kalman filter. Various methods have been used to increase the accuracy of ROMs. For example, data-driven subgrid closure models (see, e.g., \citep{Mou2021, Xie2018}) consider resolved and unresolved coherent structures associated with the truncated POD basis vectors. This approach introduces additional terms into the reduced order model, which are subsequently numerically optimized using available data. Petrov-Galerkin ROMs, e.g., derived from a least-squares approach (see, e.g., \citep{Carlberg2011}) or the Mori-Zwanzig method \citep{Parish2020} incorporate additional time-varying test basis vectors that need to be evaluated at each time step, as well. Both approaches contribute to the accuracy and stability of the reduced order model. We here employ a simple yet effective data-driven optimization method that does not introduce additional terms requiring evaluation at each time step, thereby reducing the computational effort. This choice is motivated by our objective of providing a real-time capable observer, where a computationally efficient reduced order model is needed.} 

\CommentMark{We use a greedy optimization method that finds optimal measurement locations and} ensures the number of required measurement locations to be small. We will see that the derived reduced order models recover the original CFD result for short times only (about one period corresponding to a single blade passage). In contrast, the extended Kalman filter provides reliable estimates for long times (e.g., 200 periods, see Section~\ref{sec:results_Observer}).

Section \ref{sec:Model_system} shows the numerical setup of the underlying CFD simulation. The model order reduction methods for the velocity and pressure field are presented in Section \ref{sec:vROM} and \ref{sec:pROM}, respectively. We introduce error measures to evaluate the quality of the ROMs in Section \ref{sec:Error_Eval}. The state estimation problem is solved in 
Section \ref{sec:ObserverDesign}. We evaluate the results in Section \ref{sec:results}. A brief conclusion and an outlook are stated in Section \ref{sec:conclusion_and_outlook}.

\section{Model system}\label{sec:Model_system}
We perform a flow simulation of a realistic representation of the impeller-volute interaction 
and the corresponding flow structures of a radial pump with a low specific speed ($n_s = 12\frac{1}{\text{min}}$). 
The incompressible Navier-Stokes equations read
\begin{subequations}\label{eq:NSE}
	\begin{eqnarray}
	        \frac{\partial u}{\partial t}&=&-(u\cdot \nabla)u +\nu\Delta u -\nabla p,
	        \label{eq:NSE3}
                \\
		\nabla \cdot u&=&\ 0,
		\label{eq:NSE4}
	\end{eqnarray}
\end{subequations}
where $u$ is the velocity, $p$ is the pressure and $\nu$ is the kinematic viscosity. After Reynolds-averaging and employing an eddy-viscosity turbulence model, we obtain the Reynolds-averaged Navier-Stokes equations
\begin{subequations}\label{eq:NSERANS}
        \begin{eqnarray}
		\frac{\partial \hat{u}}{\partial t}&=&-(\hat{u}\cdot \nabla)\hat{u} + \nabla [(\nu + \nu_t)\nabla\cdot \hat{u}] -\nabla \hat{p},
		\label{eq:NSE5}
		\\
		\nabla \cdot \hat{u}&=&\ 0.
		\label{eq:NSE6}
	\end{eqnarray}
\end{subequations}
In~\eqref{eq:NSERANS}, $\hat{u}$ and $\hat{p}$ denote Reynolds-averages, and $\nu_t$ is the eddy viscosity.
We use~\eqref{eq:NSERANS} for the actual flow simulation. 
In contrast, \eqref{eq:NSE} are used in the model reduction 
(specifically, in the Galerkin projections in Sections~\ref{sec:Galerkin_Projection} and~\ref{subsec:PressureROMbasedOnPressureAndVelocityModes}). 
Using~\eqref{eq:NSE} instead of~\eqref{eq:NSERANS} results in a reduced order model that, while being simpler, reproduces the original simulation results well and with a controllable error (see Section~\ref{sec:results}).

We do not distinguish $\hat{u}$ and $\hat{p}$ from $u$ and $p$ in the remainder of the text. All simulation data is obtained with~\eqref{eq:NSERANS}, while all analytical calculation such as the Galerkin projections are carried out with~\eqref{eq:NSE}.

\begin{figure*}[t!]
	\begin{minipage}[b]{0.47\textwidth}
		\centering
		\def\svgwidth{\columnwidth}
		\includegraphics[width=1\textwidth]{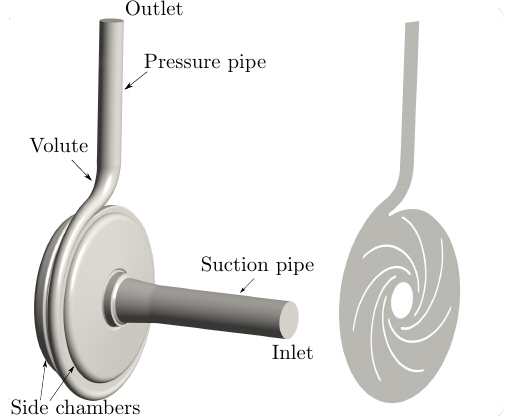}
		\captionsetup{width=\linewidth}
		\caption{Computational domain of the 3D centrifugal pump (left) and 2D axial section of the 3D pump (right).} 
		\label{fig:testcase}
	\end{minipage}
	\hfill
	\begin{minipage}[b]{0.52\textwidth}
		\centering
		\def\svgwidth[width=1\textwidth]{\columnwidth}
		\includegraphics{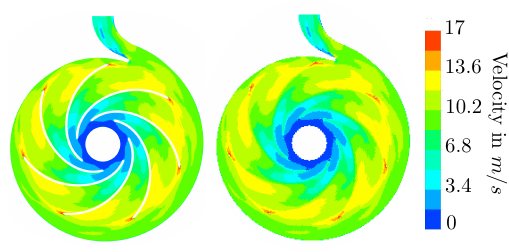}
		\captionsetup{width=\linewidth}
		\caption{Instantaneous velocity field in a body fitted, moving grid (left) and interpolated velocity field in a fixed uniform grid with artificial velocity enforced inside the blade domain (right). }
		\label{fig:Usnapshot}
	\end{minipage}
\end{figure*}

In the flow simulations, the computational domain consists of an impeller with seven blades, a spiral volute, side chambers, and the suction and pressure pipe (see Figure \ref{fig:testcase}). Body fitted, block-structured hexahedral grids with 1.8 million cells are used.
\begin{table}[b!]
\centering
\def\svgwidth{\columnwidth}
\caption{Summary of the numerical setup}
\begin{tabular}{|m{0.3\textwidth}|m{0.35\textwidth}|}
\hline
\textbf{Setup} & \textbf{Sliding grid} \\ 
\hline
CFD software & Foam-Extend Version 4.0   \\
\hline
Fluid properties &Water  \\
\hline
Solver & pimpleDyMFoam \\
\hline
Turbulence model &  SST with automatic wall function~\citep{Menter2003}\\
\hline
Time step & $1^{\circ}$ impeller rotation per time step \\ 
\hline
Pressure-velocity coupling  &  PIMPLE Algorithm \\
\hline
Time discretization & Second-order backward Euler \\
\hline
Convective discretization &  Second-order upwind TVD~\citep{WarmingandBeam1976} \\
\hline
Discretization($k$ and $\omega$) &  TVD scheme of van Leer~\citep{vanLeer1979} \\
\hline
Convergence criterion &  Nonlinear Residual $ < 10^{-5}$    \\
\hline
\textbf{Boundary conditions}  &  \\
\hline
Inlet & Velocity\\
\hline
Outlet & Static pressure \\
\hline
Angular Velocity &  $151.84\,\text{s}^{-1}$ \\
\hline
Interface & GGI \\
\hline
\end{tabular}
\label{table:numerical setup}
\end{table}
A Dirichlet inlet boundary condition is set for velocity at the nominal operating point ($ u =  2.12 \,\text{ms}^{-1}$) together with a Neumann (zero-gradient) condition for static pressure. At the outlet, Neumann boundary conditions are set for velocity (zero-gradient) and a Dirichlet condition for static pressure. The CFD simulation is conducted with OpenFOAM and the \textit{pimpleDyMFoam} solver, which combines a \textit{SIMPLE}~\citep{Patankar1972} and a \textit{PISO}~\citep{Issa1986} algorithm with moving mesh capabilities for unsteady flows. For pressure-velocity coupling, the approach from Rhie and Chow~\citep{Rhie1983} is employed. The $k$ $-$ $\omega$ $SST$ eddy viscosity turbulence model~\citep{Menter2003} is used due to its wide use for pump flow simulations in combination with automatic wall functions.
A summary of the numerical setup is given in Table \ref{table:numerical setup}. 
The convergence of the simulation is evaluated by the nonlinear and dimensionless residuals of each equation, which have to be reduced below a value of $10^{-5}$ at each time step. In addition, statistical convergence is also ensured, i.e., the change of the time-averaged characteristics (head and inner efficiency) is $<1\%$ between successive revolutions. The investigated pump model and its numerical setup are described in detail by Limbach and Skoda~\citep{Limbach2017}. We present a summary here and note that Limbach and Skoda~\citep{Limbach2017} used the commercial CFD solver ANSYS CFX 18.0. 
In contrast, we here use the open-source computational mechanics software OpenFOAM~\citep{Weller1998}.
We use the branch foam-extend 4.0, owing to the confidence we gained from the previous studies on radial pump flows~\citep{Casimir2020, Hundshagen2020}.
We compute a 3D CFD solution for the described pump model and extract a 2D axial section at the mid-span of the impeller at the nominal operating point (see Figure \ref{fig:testcase}).   
We use a 2D axial section, because it is the purpose of the present paper to demonstrate a flow field estimation in realtime is possible in principle with reduced models. 
While more technically involved, we expect the extension to the 3D case to be straight forward once the appropriate methods have been established. 
Flow fields are interpolated to a fixed cartesian grid containing all the time-variant grid solutions to simplify the model reduction steps. 
The impeller solid domain is enforced with an a posteriori approximation of interpolated values from the surrounding flow fields. This interpolation uses an algorithm based on a penalized least squares method to smooth the values between the suction and pressure side of the impeller blades~(see, e.g., \citep{Garcia2010}). In Figure \ref{fig:Usnapshot}, an example for such an interpolation from the body-fitted moving grid to the fixed cartesian grid is shown. As a result of this interpolation, the flow fields that are used for the model order reduction contain no moving structures or moving grids. This combined solid-fluid domain consideration allows the model order reduction to be carried out in a fixed grid, even for moving boundary problems.

\section{Reduced Order Model: Velocity }\label{sec:vROM}
The reconstruction of the velocity and pressure field with reduced order models requires two steps. First, a projection-based reduced order velocity model is derived. We compute spatial orthonormal basis functions, so-called POD modes, using the proper orthogonal decomposition of snapshot data with the method of snapshots~\citep{Sirovich1987a} for this purpose. Subsequently, we reduce the incompressible Navier-Stokes equations \eqref{eq:NSE} with a Galerkin projection, which results in a set of ordinary differential equations~(see, e.g., \citep{Deane1991}).

\subsection{Proper Orthogonal Decomposition}\label{sec:POD}
Simulating the spatially and temporally resolved velocity $u:\Omega \times \mathbb{R} \rightarrow\mathbb{R}^d$ of an incompressible flow on the spatial domain $\Omega\subset\mathbb{R}^d$  results in $u(x_n,t_m)$ for every discrete timestep $t_m$, $m=1, \hdots, M$ and cell $x_n$, $n=1,\hdots,N_{\text{grid}}$, on the discrete grid. 
We split $u(x_n, t_m)$ into its time-averaged contribution $\bar{u}(x_n)$ and time-variant contribution $\tilde{u}(x_n, t_m)$
\begin{align}\label{eq:Velocity_split}
	u(x_n, t_m) &= \bar{u}(x_n)+\tilde{u}(x_n,t_m),
\end{align}
\begin{equation*}
\begin{aligned}
	\bar{u}(x_n) &= \frac{1}{M}\sum\limits_{m=1}^M u(x_n,t_m),
\end{aligned}
\end{equation*}
and collect $\tilde{u}(x_n, t_m)$ in
\begin{align}
	\tilde{U} &= 
		\left[\begin{matrix}
			\tilde{u}(x_1, t_1) & \hdots & \tilde{u}(x_1, t_M)\\
			\vdots  & \ddots & \vdots\\
			\tilde{u}(x_{N_{\text{grid}}}, t_1) & \hdots & \tilde{u}(x_{N_{\text{grid}}}, t_M)\\ 
		\end{matrix}\right]\in\mathbb{R}^{N\times M},
	\label{eq:Uv}
\end{align}
where $N=dN_{\text{grid}}$.
For all simulations carried out here, $M<N_\text{grid}$ and $\text{rank}\,\tilde{U}= M$ hold.
The columns of the matrix $\Phi \in\mathbb{R}^{N\times M}$ that results from a thin singular value decomposition
\begin{align*}\label{eq:POD}
	\tilde{U}=\Phi \Sigma V^T,
\end{align*}
form a basis for the column space of $\tilde{U}$~(see, e.g., \citep{Golub2013}). Consequently, every column of $\tilde{U}$, and any linear combination of these columns, can be expressed as a linear combination of the columns
\begin{equation*}\label{eq:VeloModes}
\begin{aligned}
	\Phi_k\in\mathbb{R}^N, k= 1, \dots, M,
\end{aligned}
\end{equation*}
of $\Phi$.
Equivalently, there exist, for every column $m$ of $\tilde{U}$ in \eqref{eq:Uv}, coefficients $a_i(t_m)$, $i= 1, ..., M$, such that
\begin{equation}\label{eq:ExpansionInPhi}
	\begin{aligned}
		\begin{pmatrix}
			\tilde{u}(x_1, t_m) \\ \vdots \\ \tilde{u}(x_{N_{\text{grid}}}, t_m)
		\end{pmatrix}
		=  \sum_{i= 1}^M \Phi_i a_i(t_m)= \Phi a(t_m).
	\end{aligned}
\end{equation}
We refer to $\Phi_k=(\phi_k(x_1)^T, \dots, \phi_k(x_{N_{\text{grid}}})^T)^T$ or its components $\phi_k\in\mathbb{R}^d$ as POD modes. Rewriting 
\eqref{eq:ExpansionInPhi} in components yields the desired representation 
\begin{equation}
	\begin{aligned}
		u(x_n,t_m) = \bar{u}(x_n)  + \sum_{i=1}^M \phi_i(x_n) a_i(t_m),\label{eq:vPOD_sum}
\end{aligned}	
\end{equation}
of the flow field with its separation into spatial dependencies in $\phi_i(x_n)$ and temporal dependencies $a_i(t_m)$.  
It is the central idea of the model reduction methods used here to truncate the sum in \eqref{eq:vPOD_sum} and to retain only the most important contributions. Technically, this can be achieved by ordering the columns in $\Phi$ and the singular values $\sigma_i>0$, $i,\hdots, M$ in $\Sigma$ such that $\sigma_1\ge \sigma_2\ge \dots\ge\sigma_M$, 
and disregarding the modes $\phi_i$ for all $i>R$ for some $R< M$. This yields the approximation
	\begin{align}
		u(x_n,t_m) 
		&\approx 
		\bar{u}(x_n)  + \sum_{i=1}^R \phi_i(x_n) a_i(t_m),
		\label{eq:velo_approx}
	\end{align}
for~\eqref{eq:ExpansionInPhi} and~\eqref{eq:vPOD_sum}. 
We can control the truncation error by choosing $R$ such that 
\begin{align}\label{eq:velocity_truncationError}
	\mathcal{E}_{\text{u},\text{TRU}}(R)=1-\frac{\sum_{k=1}^R \sigma_k^2}{\sum_{k=1}^M \sigma_k^2},
\end{align}
is sufficiently small. Values of $\mathcal{E}_{\text{u},\text{TRU}}(R)\approx 1\%$ are achieved with $R=16$ in Section \ref{sec:results}. 

It is convenient to treat the time-constant mean $\bar{u}$ as a $\Phi_0$ with a constant coefficient $a_0= 1$. More precisely, let
$\Phi_0= (\phi_0(x_1)^T, \cdots, \phi_0(x_{N_{\text{grid}}})^T)^T= (\bar{u}(x_1)^T, \cdots, \bar{u}(x_{N_{\text{grid}}})^T)^T$
and $a_0(t_m)= 1$ for all $m= 1, \dots, M$.
Equation~\eqref{eq:velo_approx} can then be stated in the more compact form
\begin{align}\label{eq:compact_vROM_sum}
  u(x_n, t_m)\approx \sum_{i=0}^R\phi_i(x_n)a_i(t_m).
\end{align}

\subsection{Galerkin Projection}\label{sec:Galerkin_Projection}
The desired reduced order model can be derived by substituting~\eqref{eq:velo_approx} into the Navier-Stokes equations~\eqref{eq:NSE}, projecting the resulting equations onto the POD modes $\Phi_k$, and using their orthonormality
\begin{align}\label{eq:innerProduct}
	\delta_{kl}=\langle \Phi_k,\Phi_l\rangle = \sum_{n=1}^{N_{\text{grid}}} \phi_k(x_n)\cdot\phi_l(x_n),
\end{align}
where $\langle\cdot,\cdot\rangle$ and the dot product denote the inner products in $\mathbb{R}^{N_{\text{grid}}}$ and $\mathbb{R}^d$, respectively. This yields the ROM
\begin{subequations}\label{eq:vROM}
\begin{equation}
	\begin{aligned}\label{eq:vROM_a}
		\frac{da_k(t)}{dt} = \sum_{i=1}^R\sum_{l=1}^R  a_i(t)a_l(t)Q_{kil}+\sum_{i=1}^R a_i(t)L_{ki}+C_k
	\end{aligned}
\end{equation}
for the $a_k(t)$, $k= 1, \dots, R$, where
\begin{equation}
	\begin{aligned}\label{eq:vROM_Constants}
		Q_{kil} =&-\sum_{n=1}^{N_{\text{grid}}}\phi_k(x_n)\cdot\big(\phi_i(x_n)\cdot\nabla\big)\phi_l(x_n),\\
		L_{ki} = &\sum_{n=1}^{N_{\text{grid}}}\bigg(\nu\phi_k(x_n)\cdot\displaystyle\Delta\phi_i(x_n)
		-\phi_k(x_n)\cdot\big(\bar{u}(x_n)\cdot\nabla\big)\phi_i(x_n)
		-\phi_k(x_n)\cdot\big(\phi_i(x_n)\cdot\nabla\big)\bar{u}(x_n)\bigg),\\
		C_k =& \sum_{n=1}^{N_{\text{grid}}}\bigg(-\phi_k(x_n)\cdot\big(\bar{u}(x_n)\cdot\nabla\big)\bar{u}(x_n)
		+\nu\phi_k(x_n)\cdot\Delta\bar{u}(x_n)\bigg),
	\end{aligned}
\end{equation}
\end{subequations}
for $i= 1, \dots, R$ and $l= 1, \dots, R$.
The steps that lead to~\eqref{eq:vROM} are stated in more detail in Appendix~\ref{app:VeloROM_Derivation} for completeness. Note that the differential operators in~\eqref{eq:vROM_Constants} 
need to be approximated by finite differences on the spatial grid. 

We refer to \eqref{eq:vROM} as \textit{velocity ROM} and denote the right-hand side of \eqref{eq:vROM_a} by $f(a(t))$. We solve the velocity ROM for $a_k(t)$, $k = 1,\hdots, R$ with the given initial condition $a_k(0)=\sum_{n=1}^{N_{\text{grid}}}\tilde{u}(x_n,0)\cdot\phi_k(x_n)$, $k = 1,\hdots, R$. After solving the velocity ROM for $a_k(t)$, $k = 1,\hdots, R$, the velocity field can be reconstructed with \eqref{eq:velo_approx}.

\subsection{Optimization of the velocity ROM}\label{sec:vROM_opt}
Using different discretization schemes in the ROM than in the full order model lead to less accurate results \citep{Ingimarson2022}. We address this problem with the common practice of optimizing the coefficients $Q_{kil}$, $L_{ki}$ and $C_k$ of the ROM \eqref{eq:vROM_a} by performing a fit with the original CFD simulation data~(see, e.g., \citep{Couplet2005,Cordier2009}). \CommentMark{By avoiding the inclusion of additional closure terms~(see, e.g., \citep{Mou2021, Xie2018, Baiges2015, Zucatti2021}) or time-varying test basis vectors from Petrov-Galerkin approaches~(see, e.g., \citep{Carlberg2011, Parish2020}), the complexity of the ROM is kept at a minimum.} Note that realtime capability favors a simple ROM. We determine reference values
\begin{equation}\label{eq:vTimeFunctions_SVD}
	\begin{aligned}
		a_k^{\text{POD}}(t_m)&=\sum_{n=1}^{N_{\text{grid}}}\tilde{u}(x_n,t_m)\cdot\phi_k(x_n),
	\end{aligned}
\end{equation}
for $k = 1,\hdots, R$ and  $m = 1,\hdots, M$ for this purpose and solve
\begin{equation}\label{eq:vROM_opt}
	\begin{aligned}
	&\minA_{Q_{kil}, L_{ki}, C_{k}} \sum_{k=1}^{R}\sum_{m=1}^M (a_{k}(t_m)-a^{\text{POD}}_k(t_m))^2,
	\end{aligned}
\end{equation}
with a Levenberg-Marquardt algorithm~\citep{Levenberg1944}. \CommentMark{We use the coefficients computed by \eqref{eq:vROM_Constants} as initial coefficients for the optimization.} The velocity ROM \eqref{eq:vROM} has to be solved for every iteration of the optimization algorithm. The parameters that result from \eqref{eq:vROM_opt} are denoted $Q_{kil}^{\text{opt}}$, $L_{ki}^{\text{opt}}$ and $C_k^{\text{opt}}$. The solution of the ROM \eqref{eq:vROM_a} with these coefficients is denoted $a_k^{\text{opt}}(t_m)$. The velocities $u(x_n, t_m)$ can then be approximated by $u^{\text{opt}}(x_n,t_m) $ defined by
\begin{equation}\label{eq:vROM_approx}
		 \bar{u}(x_n)  + \sum_{i=1}^R \phi_i(x_n) a^{\text{opt}}_i(t_m)
		= \sum_{i=0}^R \phi_i(x_n) a^{\text{opt}}_i(t_m)
\end{equation}
where $a_0^{\text{opt}}(t_m)=1$ for all $m=1,\hdots,M$.

\section{Reduced Order Model: Pressure}\label{sec:pROM}
We use the Pressure-Poisson equation, which is obtained from the divergence of the momentum equation of the Navier-Stokes equations \eqref{eq:NSE3}
\begin{align}\label{eq:Pressure_Poisson_Eq}
	\Delta p = - \nabla\cdot ((u\cdot \nabla)u),
\end{align}
to derive a reduced pressure model. This equation can be used to compute the pressure $p:\Omega \times \mathbb{R} \rightarrow\mathbb{R}$ from the velocity $u$~(see, e.g., \citep{Noack2005}). We describe how to replace the partial differential equation \eqref{eq:Pressure_Poisson_Eq} by a reduced order pressure model in this section.

\subsection{Pressure ROM based on velocity modes}\label{sec:pROM_pModes}
Substituting $u^{\text{opt}}(x_n, t_m)$ defined in~\eqref{eq:vROM_approx} into \eqref{eq:Pressure_Poisson_Eq} 
and separating the spatial and temporal contributions results in
\begin{align}
	\Delta p(x_n,t_m) &=\sum_{i=0}^R\sum_{l=0}^Ra^{\text{opt}}_i(t_m)a^{\text{opt}}_l(t_m) w_{\text{Q},il}(x_n),\label{eq:Delta_vROM}\\
	w_{\text{Q},il}(x_n) &= - \nabla \cdot ((\phi_i(x_n)\cdot \nabla)\phi_l(x_n)).\nonumber
\end{align}
We seek $p_{\text{Q},il}(x_n)$ such that
\begin{align}\label{eq:v-pROM}
	p(x_n,t_m) &=\sum_{i=0}^R\sum_{l=0}^Ra^{\text{opt}}_i(t_m)a^{\text{opt}}_l(t_m) p_{\text{Q},il}(x_n),
\end{align}
respects~\eqref{eq:Delta_vROM}.
Differentiating \eqref{eq:v-pROM} and comparing coefficients to \eqref{eq:Delta_vROM} yields
\begin{align}\label{eq:v-pROM_PDE}
  \Delta p_{\text{Q},il}(x_n)= w_{\text{Q},il}(x_n)= -\nabla \cdot((\phi_i(x_n) \cdot \nabla)\phi_l(x_n)),
\end{align}
$i,l=0,\hdots,R$.
This partial differential equation only needs to be solved once for $p_{\text{Q},il}(x_n)$, $n=1,\hdots,N_{\text{grid}}$ after computing the modes $\Phi_k$. Once the $p_{\text{Q},il}(x_n)$ have been determined, the pressure field can be evaluated with the solution of the velocity ROM $a^{\text{opt}}(t_m)$ according to \eqref{eq:v-pROM}. Following~\citep{Noack2005}, we refer to \eqref{eq:v-pROM} as the {\it pressure ROM based on velocity modes}. It is convenient to collect the $p_{\text{Q},il}(x_n)$ in $P_{\text{Q}}\in\mathbb{R}^{N_{\text{grid}}\times (R+1)\times (R+1)}$.

\subsection{Pressure ROM based on pressure and velocity modes}\label{subsec:PressureROMbasedOnPressureAndVelocityModes}
\CommentMark{The fundamental steps outlined in Section \ref{sec:pROM_pModes} are used to formulate a reduced order model for pressure, incorporating an additional reduction in its dimensions.} In addition to velocity data, the CFD simulation yields spatially and temporally resolved pressure fields $P\in\mathbb{R}^{N_{\text{grid}}\times M}$. We use these pressure fields to compute additional pressure POD modes and to further reduce the size of the pressure ROM. Since $P_Q$ in the pressure ROM based on velocity modes consists of $N_{\text{grid}}\cdot (R+1)^2$ coefficients, its dimension can become prohibitively large. We reduce the pressure ROM to a size of $R_{\text{p}}\cdot (R+1)^2$, where $R_{\text{p}} \ll N_{\text{grid}}$ denotes the number of reduced pressure POD modes, which constitute the new basis vectors for a \textit{pressure ROM based on pressure and velocity modes}. 

Analogously to \eqref{eq:Velocity_split}, we split up $p(x_n,t_m)$ into its time-averaged contribution $\bar{p}(x_n)$ and time-variant contribution $\tilde{p}(x_n,t_m)$
\begin{equation*}
\begin{aligned}
	p(x_n, t_m) &= \bar{p}(x_n)+\tilde{p}(x_n,t_m),\\
\end{aligned}
\end{equation*}
collect all $\tilde{p}(x_n, t_m)$ in $\tilde{P}\in\mathbb{R}^{N_{\text{grid}}\times M}$, and perform a singular value decomposition with $\tilde{P}$. This yields the pressure POD modes $\Psi\in\mathbb{R}^{N_{\text{grid}}\times M}$, $\text{rank}\,\tilde{P}= M$, and singular values $\tau_1\ge \tau_2\ge \dots\ge\tau_M$. 
Let $\psi_k\in\mathbb{R}$ be defined by $\Psi_k= (\psi_k(x_1),\hdots , \psi_k(x_{N_{\text{grid}}}))^T$, where $\Psi_k$ refers to the $k$-th column of $\Psi$. 
We control the truncation error by choosing $R_{\text{p}}$ such that 
\begin{align}\label{eq:pressure_truncationError}
	\mathcal{E}_{\text{p},\text{TRU}}(R_{\text{p}})=1-\frac{\sum_{k=1}^{R_{\text{p}}} \tau_k^2}{\sum_{k=1}^M \tau_k^2}
\end{align}
is sufficiently small. 
This yields the approximation 
\begin{equation}
	\begin{aligned}\label{eq:pPOD_Approx}
		p(x_n, t_m) \approx \bar{p}(x_n)+  \sum_{i=1}^{R_{\text{p}}} \psi_i(x_n) b_i(t_m).
	\end{aligned}	
\end{equation}
The coefficients $b_i(t_m)$ result from the projection described in Appendix~\ref{app:PressureROM_Derivation}.
The projection yields a set of $R_{\text{p}}$ algebraic equations 
\begin{subequations}\label{eq:vp-pROM}
\begin{equation}\label{eq:vp-pROM_b}
\begin{aligned}
	b_k(t_m)=\sum_{i=0}^R\sum_{l=0}^Ra^{\text{opt}}_i(t_m)a^{\text{opt}}_l(t_m) Q_{\text{p},kil}+C_{\text{p},k},
\end{aligned}
\end{equation}
$k=1, \hdots R_{\text{p}}$ with
\begin{equation}\label{eq:vp-pROM_coeff}
	\begin{aligned}
		Q_{\text{p},kil} &= \sum_{n=1}^{N_{\text{grid}}}\psi_k(x_n) p_{\text{Q},il}(x_n),\\
		C_{\text{p},k} &= -\sum_{n=1}^{N_{\text{grid}}}\psi_k(x_n)\bar{p}(x_n),
	\end{aligned}
\end{equation}
\end{subequations}$k=1,\hdots,R_{\text{p}}$, $i=0,\hdots,R$ and $l=0,\hdots,R$.
We collect the coefficients $Q_{\text{p},kil}$ and $C_{\text{p},k}$ in $Q_{\text{p}}\in\mathbb{R}^{R_{\text{p}}\times(R+1)\times(R+1)}$ and $C_{\text{p}}\in\mathbb{R}^{R_{\text{p}}}$, respectively. 

We refer to the resulting ROM \eqref{eq:vp-pROM_b} with parameters~\eqref{eq:vp-pROM_coeff} as \textit{pressure ROM based on pressure and velocity modes}. Once \eqref{eq:vROM} has been solved and the solution $a^{\text{opt}}(t_m)$ is known, the computation of \eqref{eq:vp-pROM} requires negligible computation time, since \eqref{eq:vp-pROM} is a set of algebraic equations. 


\subsection{Optimization of the pressure ROM}\label{sec:pROM_opt}
Analogously to \eqref{eq:vROM}, we optimize the coefficients of the pressure ROM \eqref{eq:vp-pROM_b}. We determine reference values
\begin{align}\label{eq:pTimeFunctions_SVD}
        b_k^{\text{POD}}(t_m)=\sum_{n=1}^{N_{\text{grid}}}\tilde{p}(x_n,t_m)\psi_k(x_n),
\end{align}
for $k=1,\hdots, R_{\text{p}}$ and $m=1,\hdots, M$ for this purpose and solve
\begin{equation}
\begin{aligned}\label{eq:vp-pROM_opt}
	&\minA_{Q_{\text{p},kil}, C_{\text{p},k}} \sum_{k=1}^{R_{\text{p}}}\sum_{m=1}^M(b_{k}(t_m)-b^{\text{POD}}_k(t_m))^2.
\end{aligned}
\end{equation}
The parameters that result from \eqref{eq:vp-pROM_opt} are denoted $Q_{\text{p},kil}^{\text{opt}}$ and $C_{\text{p},k}^{\text{opt}}$. The pressure ROM \eqref{eq:vp-pROM} has to be evaluated for every iteration of the optimization algorithm. The solution of the pressure ROM \eqref{eq:vp-pROM_b} with these coefficients is denoted $b_{k}^{\text{opt}}(t_m)$. 
The pressure can then be approximated by $p^{\text{opt}}(x_n,t_m)$ defined by
\begin{equation}
	\begin{aligned}\label{eq:pROM_Approx}
		 \bar{p}(x_n)  + \sum_{i=1}^{R_{\text{p}}} \psi_i(x_n)b^{\text{opt}}_i(t_m).
	\end{aligned}	
\end{equation}

\section{Error Evaluation}\label{sec:Error_Eval}
For the following steps, we use the 2-norm induced by the scalar product~\eqref{eq:innerProduct}. By construction, the squared 2-norm of the difference of the velocity snapshots from the original CFD and the velocities recovered from the \CommentMark{projection onto the POD subspace} equals the sum of singular values ignored in the truncation
\begin{equation}
        \begin{aligned}\label{eq:trunc_Error_L2}
	\sum_{n=1}^{N_{\text{grid}}}\sum_{m=1}^M\bigg(\big( \tilde{u}(x_n,t_m)-\sum_{k=1}^R\phi_k(x_n)a_k^{\text{POD}}(t_m)\big)\cdot
	\big( \tilde{u}(x_n,t_m)-\sum_{k=1}^R\phi_k(x_n)a_k^{\text{POD}}(t_m)\big)\bigg)
	= \sum_{k=R+1}^M\sigma_k^2.
	\end{aligned}
\end{equation} 
This error~\eqref{eq:trunc_Error_L2} is a lower bound for any ROM that approximates $a_k^{\text{POD}}$. 
We show in Appendix~\ref{app:ROM_errors} that the ROM from Section~\ref{sec:vROM_opt} for $a_k^{\text{opt}}$ results in the error
\begin{equation}
	\begin{aligned}\label{eq:ROM_Error_L2}
			\sum_{n=1}^{N_{\text{grid}}}\sum_{m=1}^M\bigg(\big( \tilde{u}(x_n,&t_m)-\sum_{k=1}^R\phi_k(x_n)a_k^{\text{opt}}(t_m)\big)\cdot
        	\big( \tilde{u}(x_n,t_m)-\sum_{k=1}^R\phi_k(x_n)a_k^{\text{opt}}(t_m)\big)\bigg)\\
			&= \sum_{k=R+1}^M\sigma_k^2+\sum_{k=1}^R\sum_{m=1}^M(a_k^{\text{POD}}(t_m)-a_k^{\text{opt}}(t_m))^2,
	\end{aligned}
\end{equation}which amounts to~\eqref{eq:trunc_Error_L2} and an additional term as expected.
The additional term only depends on the time coefficients and will turn out to be small in Section~\ref{sec:results_ROM_Velocity}, as expected after the optimization in \eqref{eq:vROM_opt}.
We define the overall error $\mathcal{E}_{\text{u},\text{Total}}(R)$ as \eqref{eq:ROM_Error_L2} normalized by the sum of the singular values, 
i.e, $\mathcal{E}_{\text{u},\text{Total}}(R) =$
\begin{equation*}
	\frac{\sum_{k=R+1}^M\sigma_k^2+\sum_{k=1}^R\sum_{m=1}^M(a_k^{\text{POD}}(t_m)-a_k^{\text{opt}}(t_m))^2}{\sum_{k=1}^M\sigma_k^2},
\end{equation*} 
thus ensuring consistency with the truncation error \eqref{eq:velocity_truncationError}. 
Substituting \eqref{eq:velocity_truncationError} yields 
\begin{align*}
	\mathcal{E}_{\text{u},\text{Total}}(R)=\mathcal{E}_{\text{u},\text{TRU}}(R)+\mathcal{E}_{\text{u},\text{ROM}}(R),
\end{align*}
with
\begin{align*}
    \mathcal{E}_{\text{u},\text{ROM}}(R)=\frac{\sum_{k=1}^R\sum_{m=1}^M(a_k^{\text{POD}}(t_m)-a_k^{\text{opt}}(t_m))^2}{\sum_{k=1}^M\sigma_k^2}.
\end{align*}
The corresponding calculations for the pressure ROM yield 
\begin{align*}
	\mathcal{E}_{\text{p},\text{Total}}(R_{\text{p}})=\mathcal{E}_{\text{p},\text{TRU}}(R_{\text{p}})+\mathcal{E}_{\text{p},\text{ROM}}(R_{\text{p}}),
\end{align*}
\begin{align*}
    \mathcal{E}_{\text{p},\text{ROM}}(R_{\text{p}})=\frac{\sum_{k=1}^{R_{\text{p}}}\sum_{m=1}^M(b_k^{\text{POD}}(t_m)-b_k^{\text{opt}}(t_m))^2}{\sum_{k=1}^M\tau_k^2}.
\end{align*}
In addition to the errors explained so far, we report the resulting normalized and averaged velocity and pressure errors
\begin{equation}
	\begin{aligned}\label{eq:velocity_Error}
		\mathcal{E}_{\text{u}, \text{REC}}
		=\frac{1}{dN_{\text{grid}}M}\sum_{n=1}^{N_{\text{grid}}}\sum_{m=1}^M \frac{||u(x_n,t_m)-u^{\text{opt}}(x_n,t_m)||_2}{u_{\text{ref}}},
	\end{aligned}
\end{equation}
\begin{align}\label{eq:pressure_Error}
    \mathcal{E}_{\text{p}, \text{REC}}=\frac{1}{N_{\text{grid}}M}\sum_{n=1}^{N_{\text{grid}}}\sum_{m=1}^M \frac{|p(x_n,t_m)-p^{\text{opt}}(x_n,t_m)|}{p_{\text{ref}}},
\end{align}
for the reconstructed velocity and pressure fields, where $u_{\text{ref}}$ denotes the rotational velocity at the outer radius of the rotor, $p_{\text{ref}}$ denotes the specific pressure difference between the suction- and pressure side of the pump, and the 2-norm is the norm induced by the dot product introduced in~\eqref{eq:innerProduct}.

\section{Realtime flow and pressure field reconstruction with few measurements}\label{sec:ObserverDesign}
It is not practical to measure the entire spatially resolved velocity and pressure fields during the operation of the pump. 
We show how to reconstruct these fields with the reduced order model from very few  measurements of the flow field at selected points. 
While we restrict ourselves to simulated data in the present paper, 
the methods introduced in this section can be used for an online reconstruction of the fields in realtime~\citep{Gelb1974}.
The optimal locations for the selected measurement points are determined in Section~\ref{sec:Measurement_Positions}. 
The algorithm for the actual reconstruction of the fields, specifically an extended Kalman filter (EKF), is then introduced in Section~\ref{sec:ExtendedKalmanFilter}. 

\subsection{Determining optimal measurement positions}\label{sec:Measurement_Positions}
Let $t_m$ refer to an arbitrary but fixed point in time. 
Essentially, we want to determine a small number $N_\text{EKF}\ll N_\text{grid}$ of points $\xi_1, \dots, \xi_{N_\text{EKF}}$ among the grid points $x_n$, $n= 1, \dots, N_{\text{grid}}$ 
such that the velocity $u(x_n, t_m)$ and pressure $p(x_n, t_m)$ can be determined for all $x_n$ from $u(\xi_1, t_m), \dots, u(\xi_{N_\text{EKF}},t_m)$. 
According to~\eqref{eq:vROM_approx} the approximation of $u(\xi_j, t_m)$ with the optimized reduced order model for the velocity is given by
\begin{equation*}
  u(\xi_j, t_m)\approx \bar{u}(\xi_j)+ \underbrace{\sum_{i=1}^R \phi_i(\xi_j)\cdot a_i^{\text{opt}}(t_m)}_{\approx\tilde{u}(\xi_j, t_m)},
\end{equation*} 
where $j= 1, \dots N_\text{EKF}$ and $a_i^{\text{opt}}(t_m)$ are the optimized coefficients from Section~\ref{sec:vROM_opt}. 
The notation introduced in~\eqref{eq:ExpansionInPhi} can be used to express $\tilde{u}(\xi_j, t_m)$ in a compact form with the output equation
\begin{equation}\label{eq:OutputEquationForEKF}
	y(t_m)=\begin{bmatrix}
    \tilde{u}(\xi_1, t_m) 
    \\
    \vdots
    \\
    \tilde{u}(\xi_{N_\text{EKF}},t_m)
  \end{bmatrix}
  \approx \mathcal{C} a(t_m)
\end{equation}
where 
\begin{equation}\label{eq:OutputMatrixForEKF}
  \mathcal{C}
  =
  \begin{bmatrix}
    \phi_1(\xi_1) & \dots & \phi_R(\xi_1)
    \\
    \vdots & \ddots & \vdots 
    \\
    \phi_1(\xi_{N_\text{EKF}}) & \dots & \phi_R(\xi_{N_\text{EKF}})
  \end{bmatrix}\in\mathbb{R}^{dN_\text{EKF}\times R}
\end{equation}
replaces the full matrix $\Phi\in\mathbb{R}^{N\times M}$ in~\eqref{eq:ExpansionInPhi}.
We write $\mathcal{C}(\Xi)$, where $\Xi$ is short for $\xi_1, \dots, \xi_{N_\text{EKF}}$, whenever we need to point out that $\mathcal{C}$ has to be determined for candidate sets of measurement locations $\xi_i$.
The velocities $u(\xi_j, t_m)$ collected in~\eqref{eq:OutputEquationForEKF} are the outputs in our case. Using standard systems theory notation, we abbreviate the outputs by $y(t_m)$, which is introduced in~\eqref{eq:OutputEquationForEKF}. 

A fundamental result from systems theory states we can reconstruct the state of the system, i.e., the entire flow field in our case, from a restricted set of measured states or outputs, only if the observability matrix has full rank~(see, e.g., \citep{Gelb1974}).
The observability matrix here reads
\begin{align}\nonumber
	\mathcal{O}(\Xi, t_m) = 
	\left[\begin{matrix}
		\mathcal{C}(\Xi)\\
		\mathcal{C}(\Xi)\cdot J_f(t_m)\\
		\vdots\\
		\mathcal{C}(\Xi)\cdot J_f^{R-1}(t_m)
	\end{matrix}\right],
\end{align}
where $J_f(t_m)$ is the Jacobian matrix
\begin{align}\nonumber
	J_f(t_m) = \nabla_a f(a)|_{a(t_{m})},
\end{align}
with $f(a)=\frac{da_k(t)}{dt}$ from the velocity ROM~\eqref{eq:vROM}. Note that $\mathcal{O}(\Xi, t_m)$ is time-variant. 
 
The observability matrix $\mathcal{O}(\Xi, t_m)$ may have full rank but may at the same time be nearly singular. We therefore select measurement positions $\Xi$  such that $\mathcal{O}(\Xi, t_m)$ has full rank and a small condition number 
\begin{align}\label{eq:CondNum}
	\kappa (\Xi, t_m) = \frac{\delta_{\text{max}}(\mathcal{O}(\Xi, t_m))}{\delta_{\text{min}}(\mathcal{O}(\Xi, t_m))},
\end{align}
where $\delta_{\text{max}}(\mathcal{O}(\Xi, t_m))$ and $\delta_{\text{min}}(\mathcal{O}(\Xi, t_m))$ denote the largest and smallest singular value of $\mathcal{O}(\Xi, t_m)$, respectively. 
In order to ensure $\kappa(\Xi, t_m)$ is small for all times, we select the measurement positions $\Xi$ such that the largest $\kappa$ over time 
\begin{align}\nonumber
	\maxA_{m=1,\hdots ,M}\kappa (\Xi, t_m).
\end{align}
is minimized.
We use a greedy optimization algorithm (see Algorithm \ref{alg:EKF_GreedyAlg}) to find appropriate $\xi_i$, $i= 1, \dots, N_\text{EKF}$~\citep{Willcox2006}.
Here, the number of appropriate measurement locations has not been defined a priori but was chosen iteratively.

\begin{algorithm2e}[ht]
	\SetKwBlock{while}{while($j\leq N_{\text{\normalfont{EKF}}}$)}{end}
	\caption{Greedy algorithm for finding a set of measurement locations}\label{alg:EKF_GreedyAlg}
	\KwData{$N_{\text{grid}}$, $N_{\text{EKF}}$, $\Phi$, $J_f$, $\mathcal{X}=[x_1,\hdots,x_{N_{\text{grid}}}]$\;}
	\textbf{Initialize}\ : $\Xi = [\,]$, $j=1$\;
	\SetAlgoLined
	\while()
	{
		\ForAll{$x_n \in \mathcal{X}\setminus\Xi$}
		{
			$\kappa_{\text{max}}(x_n) = \underset{{m=1,\hdots ,M}}{\maxA}\kappa \left(\left[\Xi, x_n\right], t_m\right)$;
		}
		$\xi_j = \underset{x_n\in \mathcal{X}\setminus\Xi}\argminA \,\kappa_{\text{max}}(x_n)$;\\
		$j \leftarrow j+1$\;
	}
\end{algorithm2e}

\subsection{Extended Kalman filter}\label{sec:ExtendedKalmanFilter}
We use an extended Kalman filter (EKF) to determine the velocity time-functions $a(t_m)$ from the outputs $y(t_m)$, i.e., from information on the velocity field at the selected locations only. The entire velocity field $u(x_n,t_m)$, $n=1,\hdots, N_{\text{grid}}$, $m=1,\hdots, M$ in $\Omega$ can be determined with \eqref{eq:vROM_approx}, once the EKF has converged to $a(t_k)$ for an $t_k$ and provides $a(t_{k+1})$, $a(t_{k+2}), \dots$ from thereon. 

The EKF algorithm is given in Algorithm \ref{alg:EKF_ROM}. 
The EKF essentially predicts the value of the coefficients $a^{\text{EKF}}(t_m^-)$, where $t_m^-$ denotes the time immediately before the next measurement becomes available. 
This prediction is carried out by integrating the reduced order model. 
The measurement at the selected locations for time $t_m$, i.e., the output~\eqref{eq:OutputEquationForEKF} $y(t_m)$, is then used to correct the predicted value $a^{\text{EKF}}(t_m^-)$. The corrected value is denoted by $a^{\text{EKF}}(t_m)$. The matrices $\Theta_P$ and $\Theta_M$ denote the covariance of the prediction with the ROM and the covariance of the measurements, respectively. They must be known from a theoretical point of view but often are set to unit matrices multiplied with a scaling factor and used to tune the EKF in practical applications. 

Higher scaling factors in $\Theta_P$ model less confidence in the velocity ROM and a higher weighting of the measurements. Conversely, the model predictions are weighted more strongly if the scaling factor in $\Theta_M$ is chosen higher. The weighting results in the gain $K(t_m)$, which determines how strongly the deviation between the current measurement $y(t_m)$ and the current best model-based prediction $\mathcal{C} a^{\text{EKF}}(t_m^-)$ enters the new best estimate $a^{\text{EKF}}(t_m)$. 

The matrices $\Theta(t_m^-)$  and $\Theta(t_m)$ denote the covariance of the estimated values before and after the corrector step at time $t_m$. It is common practice to initialize an EKF with zero values, which read $a_i^{\text{EKF}}(0)= 0$, $i=1,\hdots, R$ here. 

\begin{algorithm2e}[H]
	\SetKwBlock{loop}{loop()}{end}
	\caption{Extended Kalman filter with velocity ROM}\label{alg:EKF_ROM}
	\KwData{$f(a)$, $\mathcal{C}$, $\Theta(0)$, $\Theta_P$, $\Theta_M$\;}
	\textbf{Initialize}\ : $m=1$, $a^{\text{EKF}}(0)=0$\;
	\SetAlgoLined
	\loop()
	{
		\textbf{Predictor:}\\
		measure $y(t_m)$\; 
		integrate $a^{\text{EKF}}(t_m^-)$ with ROM and init.\ cond.\ $a^{\text{EKF}}(t_{m-1})$\; 
		$J(t_m^-) = \nabla_a f(a)|_{a^{\text{EKF}}(t_{m-1})}$\;
		$\Theta(t_m^-) = J(t_m^-)\cdot \Theta(t_{m-1})\cdot J^T(t_m^-)+\Theta_P$\;
		\textbf{Corrector:}\\
		$K(t_m)=\Theta(t_m^-)\cdot \mathcal{C}^T\big[\mathcal{C}\cdot \Theta(t_m^-)\cdot \mathcal{C}^T+\Theta_M\big]^{-1}$\;
		$a^{\text{EKF}}(t_m) = a^{\text{EKF}}(t_m^-) + K(t_m)\big[ y(t_m)-\mathcal{C}\cdot a^{\text{EKF}}(t_m^-)\big]$\;
		$\Theta(t_m) = \big[I-K(t_m)\cdot \mathcal{C}\big]\Theta(t_m^-)$\;
		$m\leftarrow m+1$\;
	}
\end{algorithm2e}

\section{Results}\label{sec:results}
We investigate the results of the projection based ROMs in terms of the velocity and pressure field reconstruction. Additionally, we will show the resulting estimations of the velocity and pressure time-variant coefficients from observing the simulated system with the extended Kalman filter from chapter \ref{sec:ObserverDesign}.

\subsection{Results: Reduced Order Models}\label{sec:results_ROM_Velocity}
\begin{figure*}[t!]
	\begin{minipage}[b]{0.48\textwidth}
		\centering
		\def\svgwidth{\columnwidth}
		\includegraphics[width=1.05\textwidth]{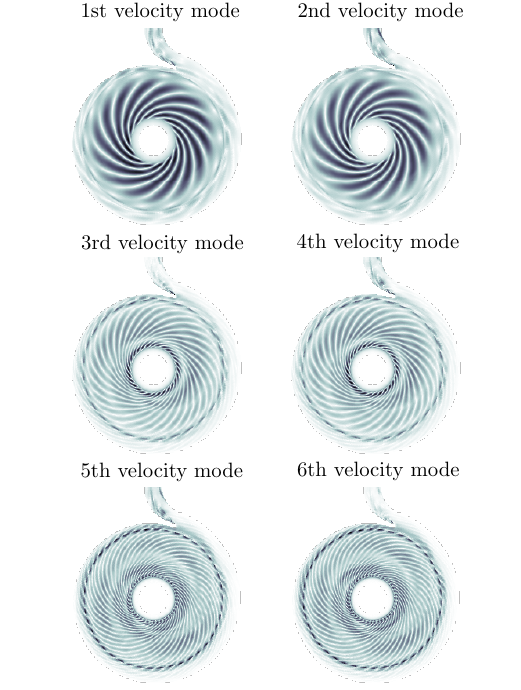}
		\captionsetup{width=\linewidth}
		\caption{Magnitude of the velocity modes $\Phi_i$, $i=1,\hdots,6$.} 
		\label{fig:slidingGrid_vPOD_Moden}
	\end{minipage}
	\hfill
	\begin{minipage}[b]{0.48\textwidth}
		\centering
		\def\svgwidth{\columnwidth}
		\includegraphics[width=1.05\textwidth]{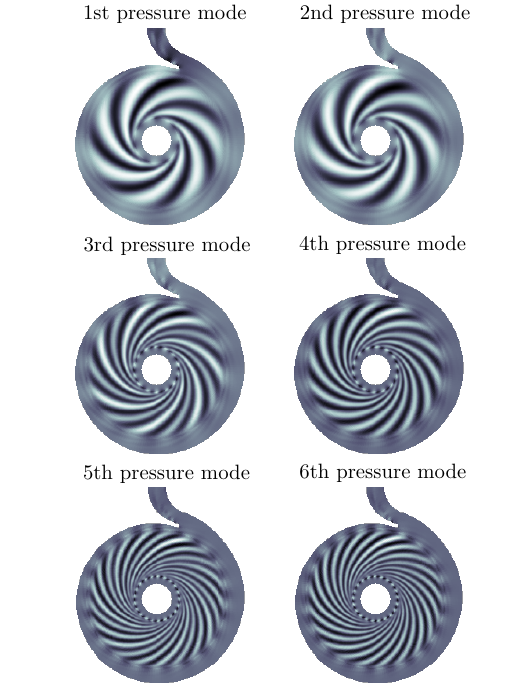}
		\captionsetup{width=\linewidth}
		\caption{Pressure modes $\Psi_i$, $i=1,\hdots,6$.}
		\label{fig:slidingGrid_pPOD_Moden}
	\end{minipage}
\end{figure*}

We use the velocity and pressure ROMs \eqref{eq:vROM} and \eqref{eq:vp-pROM} with optimized coefficients that result from \eqref{eq:vROM_opt} and \eqref{eq:vp-pROM_opt}, respectively, for the axial section of the two-dimensional velocity and pressure field of the radial pump introduced in Section \ref{sec:Model_system}. 
The CFD results obtained on a rotating grid (see Section \ref{sec:Model_system}) are interpolated onto a two-dimensional uniform cartesian grid with $236\times 262$ uniform cells in x- and y-direction, which results in $N_{\text{grid}} = 61832$ and $N=dN_{\text{grid}} = 2N_{\text{grid}}= 123664$. We capture one flow period $T_{\text{period}}$ with $52$ snapshots or $5.9\cdot 10^{-3}s$ and a sampling time of $\Delta t = 1.15\cdot 10^{-4}s$. This corresponds to one blade passage, which is used to compute the velocity $\Phi_k$ and pressure modes $\Psi_k$. The six first modes $\Phi_k$, $\Psi_k$ are shown in Figures \ref{fig:slidingGrid_vPOD_Moden} and \ref{fig:slidingGrid_pPOD_Moden} for illustration. 

\begin{figure*}[t!]
	\begin{minipage}[bt]{0.48\textwidth}
		\centering
		\def\svgwidth{\columnwidth}
		\vspace{1.5mm}
		\includegraphics[width=1\textwidth]{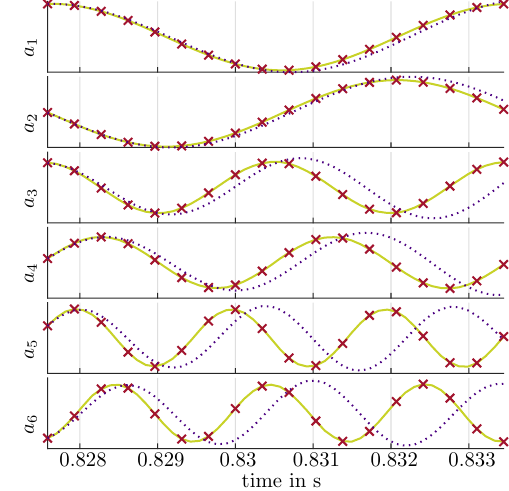}
		\captionsetup{width=\linewidth}
		\caption{Coefficients $a_i^\text{POD}$ from \eqref{eq:vTimeFunctions_SVD} (reference, green, solid), $a_i$ from the solution of \eqref{eq:vROM} (indigo, dotted) and $a_i^\text{opt}$ from the solution of \eqref{eq:vROM_a} with optimized coefficients (red crosses) for a single period $T_{\text{period}}$ and $i=1,\hdots,6$.} 
		\label{fig:slidingGrid_a}
	\end{minipage}
	\hfill
	\begin{minipage}[bt]{0.48\textwidth}
		\centering
		\def\svgwidth{\columnwidth}
		\includegraphics[width=1\textwidth]{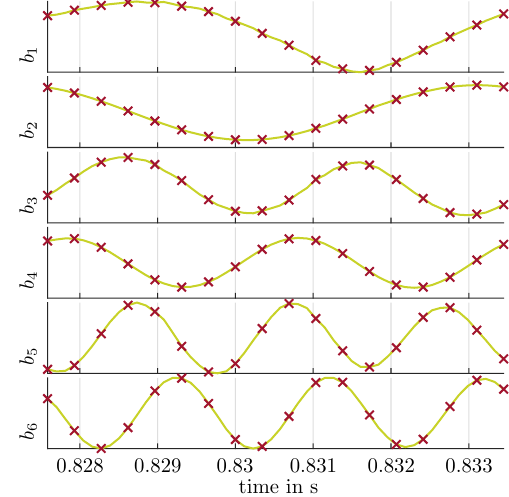}
		\captionsetup{width=\linewidth}
		\caption{Coefficients $b_i^{\text{POD}}$ from \eqref{eq:pTimeFunctions_SVD} (reference, green, solid) and $b_i^\text{opt}$ from the solution of \eqref{eq:vp-pROM_b} with optimized coefficients (red crosses) for a single period $T_{\text{period}}$ and $i=1,\hdots,6$.}
		\label{fig:slidingGrid_b}
	\end{minipage}
\end{figure*}

Table \ref{tab:truncation_errors} shows the velocity and pressure truncation errors \eqref{eq:velocity_truncationError} and \eqref{eq:pressure_truncationError} for various $R$ and $R_{\text{p}}$. We use $R=16$ basis vectors for the velocity ROM, which results in a truncation error $\mathcal{E}_{\text{u},\text{TRU}}(R)= 1.156\%$. Similarly, we use $R_{\text{p}}=16$ pressure basis vectors for the pressure ROM, which results in a truncation error $\mathcal{E}_{\text{p},\text{TRU}}(R_{\text{p}})\approx 0.3241\%$. We perform the optimizations described in Sections \ref{sec:vROM_opt} and \ref{sec:pROM_opt}, which lead to additional ROM errors of $\mathcal{E}_{\text{u},\text{ROM}}(R) \approx 0.0029\%$ and $\mathcal{E}_{\text{p},\text{ROM}}(R_{\text{p}}) \approx 1.2\times 10^{-13}\%$. The optimization error of the pressure field is much lower than for the velocity field since the reduced pressure model simply maps the results from the velocity ROM with the algebraic equation \eqref{eq:vp-pROM}. 
\begin{table}[b!]
	\centering
	\caption{Truncation errors in $\%$ for various $R$ and $R_{\text{p}}$}
	\begin{tabular*}{0.5\columnwidth}{l|lllllll}
	$R$             & \multicolumn{1}{c}{1} & \multicolumn{1}{c}{2} & \multicolumn{1}{c}{4} & \multicolumn{1}{c}{10} & \multicolumn{1}{c}{12}  & \multicolumn{1}{c}{16} \\ \hline
	$\mathcal{E}_{\text{u},\text{TRU}}$ & 57.57                & 15.615                & 6.985                 & 2.157                  & 1.683                                & 1.156                  \\ \hline
	$R_{\text{p}}$           & \multicolumn{1}{c}{1} & \multicolumn{1}{c}{2} & \multicolumn{1}{c}{4} & \multicolumn{1}{c}{10} & \multicolumn{1}{c}{12}  & \multicolumn{1}{c}{16} \\ \hline
	$\mathcal{E}_{\text{p},\text{TRU}}$ & 50.21                & 21.181                & 8.206                 & 1.191                  & 0.714                               & 0.324     
	\end{tabular*}\label{tab:truncation_errors}
	\end{table}
We evaluate the resulting velocity ROM before and after the optimization \eqref{eq:vROM_opt} in Figure \ref{fig:slidingGrid_a} for a single period $T_{\text{period}}$. The agreement of $a^{\text{POD}}(t_m)$, which represents the reference values for the time-variant coefficients with respect to $a^{\text{opt}}(t_m)$, is evident. In contrast, $a(t_m)$, i.e., the coeffients that result without the optimization \eqref{eq:vROM_opt}, show a deviation that grows with time. Although $a^{\text{opt}}(t_m)$ approximates $a^{\text{POD}}(t_m)$ well in the first period, the optimized model eventually becomes unstable. This will be further illustrated with orbits below (Figure \ref{fig:orbits}).

\begin{figure*}[b!]
	\begin{minipage}[bt]{0.48\textwidth}
		\centering
		\def\svgwidth{\columnwidth}
		\includegraphics[width=1\textwidth]{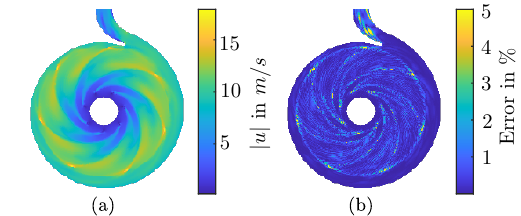}
		\captionsetup{width=\linewidth}
		\caption{(a) Approximated magnitude of the instantaneous velocity field with $a_i^{\text{opt}}(t_1)$, $i=1,\hdots,R$ and \eqref{eq:vROM_approx} and (b) relative error of the approximation with respect to the interpolated CFD result, scaled to the rotational velocity of the outer radius of the impeller $u_{\text{ref}} = 16.7\,m/s$ for the first timestep.}  
		\label{fig:slidingGrid_vError}
	\end{minipage}
	\hfill
	\begin{minipage}[bt]{0.48\textwidth}
		\centering
		\def\svgwidth{\columnwidth}
		\includegraphics[width=1\textwidth]{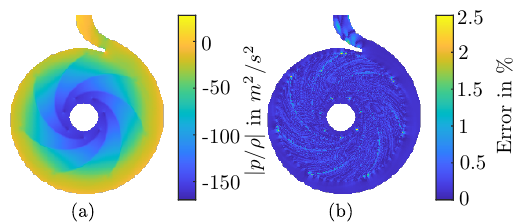}
		\captionsetup{width=\linewidth}
		\caption{(a) Approximated magnitude of the instantaneous pressure field with $b_i^{\text{opt}}(t_1)$, $i=1,\hdots,R$ and \eqref{eq:pROM_Approx} and (b) relative error of the approximation with respect to the interpolated CFD result, scaled to the referential pressure difference between suction and pressure side of the pump $p_{\text{ref}}=152.46\,m^2/s^2$ for the first timestep.}  
		\label{fig:slidingGrid_pError}
	\end{minipage}
\end{figure*}

We use the coefficients $a^{\text{opt}}(t_m)$ to reconstruct the two-dimensional velocity field with \eqref{eq:vROM_approx}. The magnitude of the resulting approximation of the velocity field, and the relative error of this approximation with respect to the original CFD results, are shown in Figure \ref{fig:slidingGrid_vError}. All values in this figure are scaled to the rotational velocity at the outer radius of the impeller $u_\text{ref} = 16.7\, m/s$. The temporally and spatially averaged deviation of the velocity field reconstruction and the original velocity field from the CFD are very small. The reconstruction error introduced in \eqref{eq:velocity_Error} amounts to $\mathcal{E}_{\text{u}, \text{REC}}=0.2933\%$. Some isolated maximum errors reach $~ 5\%$. Here, the truncation error \eqref{eq:velocity_truncationError} constitutes the largest contribution to the error of the velocity field. The optimization method \eqref{eq:vROM_opt} only introduces the additional ROM error of $\mathcal{E}_{\text{u},\text{ROM}}(R)\approx 0.0029\%$, which is three orders of magnitude smaller than the truncation error $\mathcal{E}_{\text{u},\text{TRU}}(R)= 1.156\%$.

Results for the optimized pressure ROM are shown in Figure \ref{fig:slidingGrid_b} for the same period. The comparison of $b^{\text{opt}}(t_m)$ to $b^{\text{POD}}(t_m)$ also indicates a very good agreement. Since these results are based on the results of the velocity ROM, the pressure ROM eventually becomes unstable, too. Figure \ref{fig:slidingGrid_pError} shows the reconstruction of the pressure field for the time-variant coefficients $b^{\text{opt}}(t_m)$ with \eqref{eq:pROM_Approx} and the error of this approximation with respect to the results obtained from the CFD simulation. The error is scaled to the specific pressure difference between suction and pressure side $p_{\text{ref}}=152.46$ $m^2/s^2$. The difference in the pressure fields that result from the pressure ROM and the CFD is very small again. The reconstruction error \eqref{eq:pressure_Error} amounts to $\mathcal{E}_{\text{p}, \text{REC}}=0.1337\%$. Some isolated errors reach $2.5\%$ in this case. Just as for the velocity field, the error mostly originates from the truncation error $\mathcal{E}_{\text{p},\text{TRU}}(R_{\text{p}})  \approx 0.3241\%$. The ROM error $\mathcal{E}_{\text{p},\text{ROM}}(R_{\text{p}}) \approx 1.2\times 10^{-13}\%$ is negligible, in comparison.

\subsection{Results: Extended Kalman filter}\label{sec:results_Observer}

\begin{figure}[t!]
	\centering
	\def\svgwidth{\columnwidth}
	\includegraphics{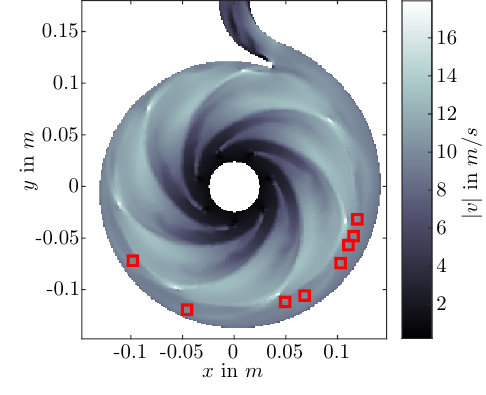}
	\captionsetup{width=\linewidth}
	\caption{Sensor locations (red squares) used for the EKF.} 
	\label{fig:slidingGrid_EKF_MeasurementLocations}
\end{figure}

It is the purpose of the extended Kalman filter to provide information about the current state of the system. Consequently, the extended Kalman filter is useful only if it is stable over many periods. We choose to analyse $200$ periods. We stress this number is arbitrary. The results presented here show that it is reasonable to assume the extended Kalman filter to be long-time stable, however. 

\begin{figure*}[b!]
	\begin{minipage}[bt]{0.48\textwidth}
		\centering
		\def\svgwidth{\columnwidth}
		\includegraphics[width=1\textwidth]{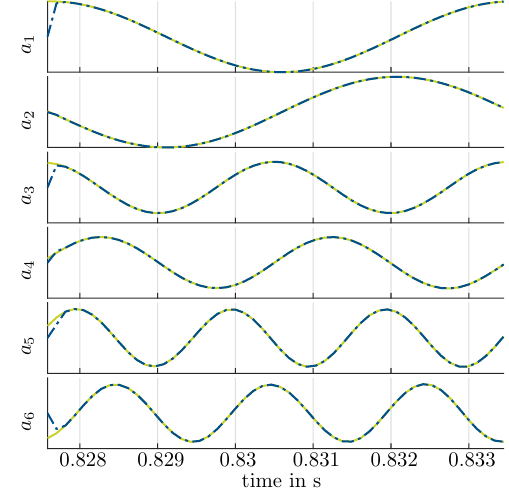}
		\captionsetup{width=\linewidth}
		\caption{Coefficients $a_i^{\text{POD}}$ from \eqref{eq:vTimeFunctions_SVD} (reference, green, solid) and coefficients estimated with the extended Kalman filter $a_i^{\text{EKF}}$  (blue, dash-dotted) for a single period $T$ and $i=1,\hdots,6$.}
		\label{fig:slidingGrid_EKF_a}
	\end{minipage}
	\hfill
	\begin{minipage}[bt]{0.48\textwidth}
		\centering
		\def\svgwidth{\columnwidth}
		\includegraphics[width=1\textwidth]{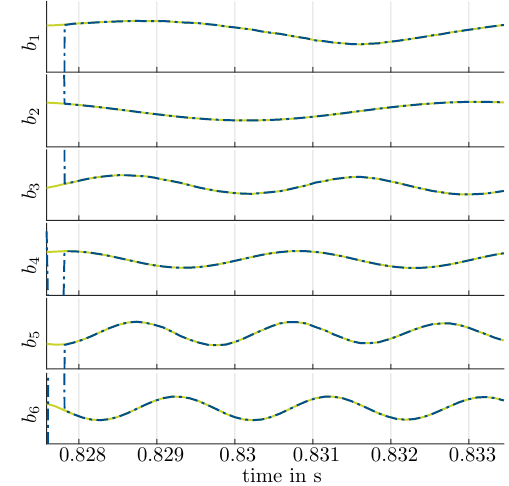}
		\captionsetup{width=\linewidth}
		\caption{Coefficients $b_i^{\text{POD}}$ from \eqref{eq:pTimeFunctions_SVD} (reference, green, solid) and coefficients estimated with the extended Kalman filter, $b_i^{\text{EKF}}$  (blue, dash-dotted) for a single period $T$ and $i=1,\hdots,6$.}
		\label{fig:slidingGrid_EKF_b}
	\end{minipage}
\end{figure*}

\begin{figure*}[t!]
	\begin{minipage}[b]{0.325\textwidth}
		\centering
		\def\svgwidth{\columnwidth}
		\includegraphics[width=1.05\textwidth]{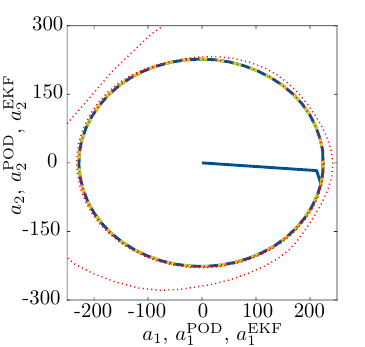}
	\end{minipage}
	\hfill
	\begin{minipage}[b]{0.325\textwidth}
		\centering
		\def\svgwidth{\columnwidth}
		\includegraphics[width=1.05\textwidth]{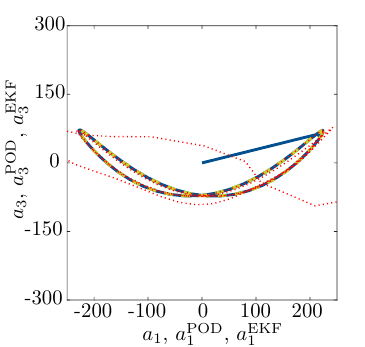}
	\end{minipage}
	\begin{minipage}[b]{0.325\textwidth}
		\centering
		\centering
		\def\svgwidth{\columnwidth}
		\includegraphics[width=1.05\textwidth]{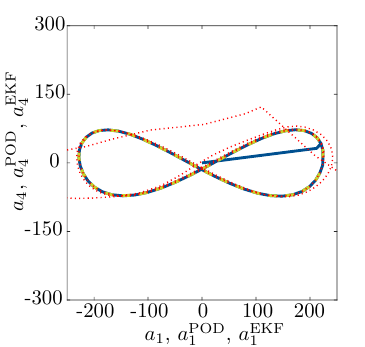}
	\end{minipage}
	\caption{Orbits showing the limit cycle:  $a_{i}^{\text{opt}}$ with optimized coefficients for two periods $T$ (red, dotted), $a_i^\text{POD}$ (green),  $a^{\text{EKF}}_{i}$ for 200 periods $T$ (blue, dashed). While the ROM becomes unstable, the observed time-variant coefficients $a_i^{\text{EKF}}$ are stable for many periods (here: 200 periods).}\label{fig:orbits}
\end{figure*}

The covariance matrices introduced in Section~\ref{sec:ExtendedKalmanFilter} 
are set to $\Theta(0)=2I_{R}$ and $\Theta_P= I_{R}$, respectively, where $I_{R}$ is the $R\times R$ unit matrix. We choose the measurement noise covariance to be $\Theta_M=I_{dN_{\text{EKF}}}$, where $I_{dN_{\text{EKF}}}\in\mathbb{R}^{dN_{\text{EKF}}\times dN_{\text{EKF}}}$ as we measure the velocity in both x- and y-coordinate direction. The initial time-variant velocity coefficients $a_i^{\text{EKF}}(0)$ are not known and are set to $a_i^{\text{EKF}}(0) = 0$, $i=1,\hdots,R$. The analysis has shown that $N_{\text{EKF}}=8$ measurement positions (see Figure \ref{fig:slidingGrid_EKF_MeasurementLocations}) are sufficient to estimate the time-variant velocity coefficients for our specific system. Fewer than $N_{\text{EKF}}=8$ positions do not result in a stable extended Kalman filter.

The estimation of the pump state with the extended Kalman filter with virtual measurements from a simulated pump are shown in Figures \ref{fig:slidingGrid_EKF_a} and \ref{fig:slidingGrid_EKF_b} for one period. The estimation converges to a stable limit cycle. This limit cycle is illustrated in Figure \ref{fig:orbits} by plotting $200$ periods. The limit cycle that results for the reference data $a^{\text{POD}}(t_m)$, which is also shown in Figure  \ref{fig:orbits}, is practically indistinguishable from the values estimated with the extended Kalman filter. It takes about $2\Delta t=2.3\times 10^{-4}$ s for the Kalman filter to converge. 

\CommentMark{The values of $K(t_m)$ increase for a few initial steps. More precisely, the Frobenius norm of the Kalman gain, denoted by $\lVert K(t_m )\rVert_F$, shows an initial value of approximately $5.75\cdot 10^3$, which increases to $2.32\cdot 10^4$ during the few time steps and does not change considerably afterwards anywhere. This increase turns out not to be significant, however, since the 2-norm of the error in the predictions projected onto the measured locations $C\cdot a^{\text{EKF}}(t_m^-)$ with regard to the measurements $y(t_m)$ indicates that the difference between $C\cdot a^{\text{EKF}}(t_m^-)$ and $y(t_m)$ is negligible. Initially, this 2-norm amounts to $1.03$, but diminishes to $5.14\cdot 10^{-6}$ after the first initial steps. Consequently, after the initial few time steps, only minor corrections of the predicted states are necessary, typically in the order of $10^{-2}$. Conversely, the corrections of the predicted states $a^{\text{EKF}}(t_m^-)$ made during the initial steps are significant, which is to be expected, as the initial values of $a^{\text{EKF}}$ are unknown and arbitrarily set to zero. Thus, it is evident that both the reduced order model and the measurements are necessary for this methodology to be effective.} 

We recall the integration of the velocity ROM, i.e., the prediction of the pump state without the extended Kalman filter, was not stable. The corresponding orbits, which are shown in Figure \ref{fig:orbits} for comparison, clearly indicate the velocity ROM itself cannot replace the extended Kalman filter.

The error introduced by the estimated coefficients $a^{\text{EKF}}(t_m)$ from the extended Kalman filter amounts to $\mathcal{E}_{\text{u},\text{ROM}}(R)= 0.0029\%$ and, thus, is negligible compared to $\mathcal{E}_{\text{u},\text{TRU}}(R)= 1.156\%$. The overall mean velocity field reconstruction error $\mathcal{E}_{\text{u},\text{REC}}(R)=0.2932\%$. The results for the pressure reconstruction with estimated states provide similar results as the velocity field estimation. The error resulting from the estimated coefficients $b^{\text{EKF}}(t_m)$ amounts to $\mathcal{E}_{\text{p},\text{ROM}}(R_{\text{p}})= 7\times 10^{-5}\%$. 
This additional error is negligible compared to $\mathcal{E}_{\text{p},\text{TRU}}(R_{\text{p}})= 0.3241\%$. The mean error of the reconstructed pressure field reads $\mathcal{E}_{\text{p}, \text{REC}}=0.1338\%$. These results are practically equal to those from the direct integration of the velocity ROM but show the estimation for over 200 periods, whereas the results for the velocity ROM only hold for a single period and become unstable afterwards.

\section{Conclusion and Outlook}\label{sec:conclusion_and_outlook}
We showed that reduced order models can be used to reconstruct the velocity and pressure field of centrifugal pumps. Reduced order models were constructed using proper orthogonal decomposition on velocity and pressure snapshots generated with URANS CFD simulations. A Galerkin projection has then transformed the Navier-Stokes and Pressure-Poisson equations to sets of ordinary differential and algebraic equations, respectively. The results of the velocity and pressure fields indicate a good reconstruction in terms of accuracy and computational effort. The evaluation of appropriate sensor placement locations with the proposed greedy algorithm led to a converging extended Kalman filter. Since evaluating the reduced order model requires much less effort in terms of computational demands than computing the CFD simulation, the extended Kalman filter can be used in realtime for online processes.

It was the purpose of the present paper to show a flow field estimation in realtime is possible in principle with reduced order models. We used a 2D axial section of a 3D CFD model for this purpose. 
Future research will focus on performing the model order reduction for three-dimensional pump geometries. 

Measurements inside a centrifugal pump pose a great challenge. To enable real world flow estimations with this technique, we will investigate if velocity measurements can be replaced by pressure measurements.

\section*{Acknowlegdement}
Funded by the Federal Ministry for Economic Affairs and Climate Action (BMWK) through the AiF (German Federation of Industrial Research Associations eV) based on a decision taken by the German Bundestag (IGF no. 20275 N) and the Deutsche Forschungsgemeinschaft (DFG, German Research Foundation) – Project-ID 422037413 – TRR 287.

\bibliographystyle{agsm}
\bibliography{bibtex}

\section*{Appendix}
\appendix
\section{Derivation of the velocity ROM}\label{app:VeloROM_Derivation}
Let $t_m\in\{1,\hdots,M\}$ be arbitrary. Assuming the Navier-Stokes equations \eqref{eq:NSE} have been solved on the spatial and temporal grid, we have
\begin{equation}
	\begin{aligned}\label{eq:AppendixA_NSE_array}
		&\left(\begin{matrix}
			\dfrac{\partial u(x_1,t_m)}{\partial t}\\
			\vdots\\
			\dfrac{\partial u(x_{N_{\text{grid}}},t_m)}{\partial t}
		\end{matrix}\right)
		= 
		-\left(\begin{matrix}
			(u(x_1,t_m)\cdot\nabla)u(x_1,t_m)\\
			\vdots\\
			(u(x_{N_{\text{grid}}},t_m)\cdot\nabla)u(x_{N_{\text{grid}}},t_m)
		\end{matrix}\right)
		+
		\left(\begin{matrix}
			\nu\Delta u(x_1,t_m)\\
			\vdots\\
			\nu\Delta u(x_{N_{\text{grid}}},t_m)
		\end{matrix}\right)
		-
		\left(\begin{matrix}
			\nabla p(x_1,t_m)\\
			\vdots\\
			\nabla p(x_{N_{\text{grid}}},t_m)
		\end{matrix}\right),
	\end{aligned}
\end{equation}
where all evaluations at points $x_n$, $t_m$ are understood to be carried out after the respective differentiations. Substituting~\eqref{eq:velo_approx} into the left-hand side of~\eqref{eq:AppendixA_NSE_array}, projecting onto $\Phi_k$ and using~\eqref{eq:innerProduct} yields
\begin{equation}
	\begin{aligned}\label{eq:AppendixA_LHS}
		\left\langle\Phi_k,\sum_{i=1}^{R}\Phi_i\dfrac{\partial a_i(t_m)}{\partial t}\right\rangle
		=\sum_{i=1}^{R}\langle\Phi_k,\Phi_i\rangle\dfrac{\partial a_i(t_m)}{\partial t}
		=\sum_{i=1}^{R}\delta_{ki}\dfrac{\partial a_i(t_m)}{\partial t}
		=\dfrac{\partial a_k(t_m)}{\partial t_m}
		=\dfrac{d a_k(t)}{d t},
	\end{aligned}
\end{equation}
for all $k=1,\hdots,R$, where we replaced the partial derivative with respect to $t$, because the coefficients $a_k$ only depend on time. Applying the same steps to the first term on the right-hand side of~\eqref{eq:AppendixA_NSE_array}, 
\begin{equation*}
	\begin{aligned}\label{eq:AppendixA_RHS1}
		&\left\langle\Phi_k,-
		\left(\begin{matrix}
			\bigg(\big(\bar{u}(x_1)  + \displaystyle\sum_{i=1}^R \phi_i(x_1) a_i(t_m)\big)\cdot\nabla\bigg)\big(\bar{u}(x_1)  + \displaystyle\sum_{l=1}^R \phi_l(x_1) a_l(t_m)\big)\\
			\vdots\\
			\bigg(\big(\bar{u}(x_{N_{\text{grid}}})  + \displaystyle\sum_{i=1}^R \phi_i(x_{N_{\text{grid}}}) a_i(t_m)\big)\cdot\nabla)\bigg)\big(\bar{u}(x_{N_{\text{grid}}})  + \displaystyle\sum_{l=1}^R \phi_l(x_{N_{\text{grid}}}) a_l(t_m)\big)
		\end{matrix}\right)\right\rangle
	\end{aligned}
\end{equation*}
\begin{equation}
	\begin{aligned}\label{eq:AppendixA_RHS1}
		&= -\sum_{n=1}^{N_{\text{grid}}}\phi_k(x_n)\cdot\bigg(\big(\bar{u}(x_n)+\sum_{i=1}^R\phi_i(x_n)a_i(t_m)\big)\cdot\nabla\bigg)\big(\bar{u}(x_n)+\sum_{l=1}^R\phi_l(x_n)a_l(t_m)\big)\\
		&= -\sum_{n=1}^{N_{\text{grid}}}\phi_k(x_n)\cdot\big(\bar{u}(x_n)\cdot\nabla\big)\bar{u}(x_n)
		-\sum_{l=1}^R a_l(t_m)\sum_{n=1}^{N_{\text{grid}}}\phi_k(x_n)\cdot\big(\bar{u}(x_n)\cdot\nabla\big)\phi_l(x_n)\\
		&\quad\,-\sum_{i=1}^R a_i(t_m)\sum_{n=1}^{N_{\text{grid}}}\phi_k(x_n)\cdot\big(\phi_i(x_n)\cdot\nabla\big)\bar{u}(x_n)
		-\sum_{i=1}^R\sum_{l=1}^R a_i(t_m) a_l(t_m)\sum_{n=1}^{N_{\text{grid}}}\phi_k(x_n)\cdot\big(\phi_i(x_n)\cdot\nabla\big)\phi_l(x_n).
	\end{aligned}
\end{equation}
The second term on the right-hand side of~\eqref{eq:AppendixA_NSE_array} can be treated analogously to give
\begin{equation}
	\begin{aligned}\label{eq:AppendixA_RHS2}
		\left\langle\Phi_k,
		\left(\begin{matrix}
			\nu\bigg(\Delta\bar{u}(x_1)+\displaystyle\sum_{i=1}^R \Delta\phi_i(x_1)a_i(t_m)\bigg)\\
			\vdots\\
			\nu\bigg(\Delta\bar{u}(x_{N_{\text{grid}}})+\displaystyle\sum_{i=1}^R \Delta\phi_i(x_{N_{\text{grid}}})a_i(t_m)\bigg)
		\end{matrix}\right)\right\rangle
		=\sum_{n=1}^{N_{\text{grid}}}\nu\phi_k(x_n)\cdot\bigg(\Delta\bar{u}(x_n)+\displaystyle\sum_{i=1}^R \Delta\phi_i(x_n)a_i(t_m)\bigg)\\
		=\sum_{n=1}^{N_{\text{grid}}}\nu\phi_k(x_n)\cdot\Delta\bar{u}(x_n)
		+\sum_{i=1}^R a_i(t_m)\sum_{n=1}^{N_{\text{grid}}}\nu\phi_k(x_n)\cdot\displaystyle\Delta\phi_i(x_n).
	\end{aligned}
\end{equation}
Equating the left-hand side~\eqref{eq:AppendixA_LHS} with the right-hand side that results from adding~\eqref{eq:AppendixA_RHS1} and~\eqref{eq:AppendixA_RHS2}, and collecting terms constant, linear, and quadratic in $a_i(t_m)$ yields~\eqref{eq:vROM}. 
The term in~\eqref{eq:AppendixA_NSE_array} that depends on the pressure gradient is usually neglected~(see, e.g., \citep{John2010}). The continuity equation \eqref{eq:NSE4} is also neglected in the ROM formulation, since the zero divergence of the velocity is already guaranteed for the CFD simulation data.

\section{Derivation of the pressure ROM based on pressure and velocity modes}\label{app:PressureROM_Derivation}
Let $t_m\in\{1,\hdots,M\}$ be arbitrary. Assuming the pressure ROM based on velocity modes \eqref{eq:v-pROM_PDE} has been solved on the spatial and temporal grid, we have
\begin{equation}
	\begin{aligned}\label{eq:v_pROM_PDE_array}
		&\left(\begin{matrix}
			p(x_1,t_m)\\
			\vdots\\
			p(x_{N_{\text{grid}}},t_m)
		\end{matrix}\right)
		= 
		\left(\begin{matrix}
			\displaystyle\sum_{i=0}^R\displaystyle\sum_{l=0}^Ra^{\text{opt}}_i(t_m)a^{\text{opt}}_l(t_m) p_{\text{Q},il}(x_1)\\
			\vdots\\
			\displaystyle\sum_{i=0}^R\displaystyle\sum_{l=0}^Ra^{\text{opt}}_i(t_m)a^{\text{opt}}_l(t_m) p_{\text{Q},il}(x_{N_{\text{grid}}})
		\end{matrix}\right).
	\end{aligned}
\end{equation}
Substituting~\eqref{eq:pPOD_Approx} into the left-hand side of~\eqref{eq:v_pROM_PDE_array}, projecting onto $\Psi_k$ and using~\eqref{eq:innerProduct} yields
\begin{equation}
	\begin{aligned}\label{eq:AppendixB_LHS}
		\left\langle\Psi_k,\left(\begin{matrix}\bar{p}(x_1)\\\vdots\\\bar{p}(x_{N_{\text{grid}}})\end{matrix}\right)+\sum_{i=1}^{R_{\text{p}}}\Psi_i b_i(t_m)\right\rangle
		&=\left\langle\Psi_k,\left(\begin{matrix}\bar{p}(x_1)\\\vdots\\\bar{p}(x_{N_{\text{grid}}})\end{matrix}\right)\right\rangle+\sum_{i=1}^{R_{\text{p}}}\langle\Psi_k,\Psi_i\rangle b_i(t_m)\\
		&=\sum_{n=1}^{N_{\text{grid}}}\psi_k(x_n)\bar{p}(x_n)+\sum_{i=1}^{R_{\text{p}}}\delta_{ki}b_i(t_m)
		=\sum_{n=1}^{N_{\text{grid}}}\psi_k(x_n)\bar{p}(x_n)+b_k(t_m),
	\end{aligned}
\end{equation}
for all $k=1,\hdots,R_{\text{p}}$. Applying the same steps to the term on the right-hand side of~\eqref{eq:v_pROM_PDE_array}, 
\begin{equation}
	\begin{aligned}\label{eq:AppendixB_RHS}
		\left\langle\Psi_k,\left(\begin{matrix}
			\displaystyle\sum_{i=0}^R\displaystyle\sum_{l=0}^Ra^{\text{opt}}_i(t_m)a^{\text{opt}}_l(t_m) p_{\text{Q},il}(x_1)\\
			\vdots\\
			\displaystyle\sum_{i=0}^R\displaystyle\sum_{l=0}^Ra^{\text{opt}}_i(t_m)a^{\text{opt}}_l(t_m) p_{\text{Q},il}(x_{N_{\text{grid}}})
		\end{matrix}\right)\right\rangle
		&= \sum_{n=1}^{N_{\text{grid}}}\psi_k(x_n)\sum_{i=0}^R\sum_{l=0}^Ra^{\text{opt}}_i(t_m)a^{\text{opt}}_l(t_m) p_{\text{Q},il}(x_n)\\
		&= \sum_{i=0}^R\sum_{l=0}^Ra^{\text{opt}}_i(t_m)a^{\text{opt}}_l(t_m) \sum_{n=1}^{N_{\text{grid}}}\psi_k(x_n) p_{\text{Q},il}(x_n).
	\end{aligned}
\end{equation}
Equating the left-hand side~\eqref{eq:AppendixB_LHS} with the right-hand side~\eqref{eq:AppendixB_RHS}, yields~\eqref{eq:vp-pROM}.

\section{Truncation and ROM errors}\label{app:ROM_errors}
The squared error of the original velocity field $u(x_n,t_m)=\bar{u}(x_n)+\tilde{u}(x_n,t_m)$ to the field approximated with the ROM $u^{\text{opt}}(x_n,t_m)=\bar{u}(x_n)+\sum_{k=1}^R\phi_k(x_n)a_k^{\text{opt}}(t_m)$ reads
\begin{equation}\label{eq:Appendix_e}
	\begin{aligned}
		&\sum_{n=1}^{N_{\text{grid}}}\sum_{m=1}^{M}\Big(\big(\tilde{u}(x_n,t_m)-\sum_{k=1}^R\phi_k(x_n)a_k^{\text{opt}}(t_m)\big)\cdot
		\big(\tilde{u}(x_n,t_m)-\sum_{k=1}^R\phi_k(x_n)a_k^{\text{opt}}(t_m)\big)\Big)\\
		&= \sum_{n=1}^{N_{\text{grid}}}\sum_{m=1}^{M}\Big(\tilde{u}(x_n,t_m)\cdot \tilde{u}(x_n,t_m)
		- 2\sum_{k=1}^R a_k^{\text{opt}}(t_m)\phi_k(x_n)\cdot\tilde{u}(x_n,t_m)\\
		&\qquad+\sum_{k=1}^R\sum_{l=1}^Ra_k^{\text{opt}}(t_m)a_l^{\text{opt}}(t_m)\phi_k(x_n)\cdot\phi_l(x_n)\Big).
	\end{aligned}
\end{equation}
Equation~\eqref{eq:Appendix_e} needs to be simplified with~\eqref{eq:vTimeFunctions_SVD} and 
\begin{subequations}
	\begin{equation}
		\sum_{n=1}^{N_{\text{grid}}}\sum_{m=1}^{M}\tilde{u}(x_n,t_m)\cdot\tilde{u}(x_n,t_m)=\sum_{m=1}^M\sum_{k=1}^Ma_k^{\text{POD}}(t_m)^2,\label{eq:app_3_utilde_apod}
	\end{equation}
	\begin{equation}
		\sum_{m=1}^M\sum_{k=1}^M a_k^{\text{POD}}(t_m)^2 =\sum_{k=1}^M \sigma_k^2,\label{eq:app_3_apod_sigma}
	\end{equation}
\end{subequations}
where~\eqref{eq:app_3_utilde_apod} follows with $\tilde{u}(x_n,t_m)= \sum_{i=1}^M \phi_i(x_n) a_i^\text{POD}(t_m)$ and the orthonormality~\eqref{eq:innerProduct}, 
and~\eqref{eq:app_3_apod_sigma} follows with \eqref{eq:app_3_utilde_apod} and because the squared Frobenius norm of $\tilde{U}$, which is equal to the left-hand side\ of~\eqref{eq:app_3_utilde_apod},
is equal to the sum of the squared singular values of $\tilde{U}$~\citep{Golub2013}.
Substituting \eqref{eq:vTimeFunctions_SVD}, \eqref{eq:app_3_utilde_apod} and \eqref{eq:app_3_apod_sigma} into~\eqref{eq:Appendix_e} yields
\begin{align*}
	&\sum_{m=1}^M\sum_{k=1}^Ma_k^{\text{POD}}(t_m)^2+\sum_{m=1}^M\sum_{k=1}^R \Big(-2 \big(a_k^{\text{POD}}(t_m)a_k^{\text{opt}}(t_m)\big)+a_k^{\text{opt}}(t_m)^2\Big)\\
	&=\sum_{m=1}^M\sum_{k=R+1}^Ma_k^{\text{POD}}(t_m)^2+\sum_{m=1}^M\sum_{k=1}^R \Big(a_k^{\text{POD}}(t_m)^2 -2 \big(a_k^{\text{POD}}(t_m)a_k^{\text{opt}}(t_m)\big)+a_k^{\text{opt}}(t_m)^2\Big)\\
	&= \sum_{k=R+1}^M\sigma_k^2+\sum_{m=1}^M\sum_{k=1}^R \Big(a_k^{\text{POD}}(t_m)^2 -2 \big(a_k^{\text{POD}}(t_m)a_k^{\text{opt}}(t_m)\big)+a_k^{\text{opt}}(t_m)^2\Big)\\
	&= \sum_{k=R+1}^M\sigma_k^2+\sum_{m=1}^M\sum_{k=1}^R \Big(a_k^{\text{POD}}(t_m)-a_k^{\text{opt}}(t_m)\Big)^2
\end{align*}
We scale the error to the squared norm of the snapshots 
\begin{align*}
\sum_{n=1}^{N_{\text{grid}}}\sum_{m=1}^{M}\tilde{u}(x_n,t_m)\cdot\tilde{u}(x_n,t_m)=\sum_{k=1}^M\sigma_k^2.
\end{align*}
The scaled total error then yields
\begin{align*}
	&\frac{\sum_{k=R+1}^M\sigma_k^2}{\sum_{k=1}^M\sigma_k^2}+\frac{\sum_{m=1}^M\sum_{k=1}^R \Big(a_k^{\text{POD}}(t_m)-a_k^{\text{opt}}(t_m)\Big)^2}{\sum_{k=1}^M\sigma_k^2}\\
	&=\underbrace{1-\frac{\sum_{k=1}^R\sigma_k^2}{\sum_{k=1}^M\sigma_k^2}}_{\mathcal{E}_{\text{TRU}}}+\underbrace{\frac{\sum_{m=1}^M\sum_{k=1}^R \Big(a_k^{\text{POD}}(t_m)-a_k^{\text{opt}}(t_m)\Big)^2}{\sum_{k=1}^M\sigma_k^2}}_{\mathcal{E}_{\text{ROM}}},
\end{align*}
where $\mathcal{E}_{\text{TRU}}$ denotes the truncation error and $\mathcal{E}_{\text{ROM}}$ an additional error induced by the model order reduction.

\typeout{col width is \the\columnwidth} 

\end{document}

%% file: nomenclature.tex
\nomenclature[$u$]{$u \in \mathbb{R}^d$}{velocity}
\nomenclature[$u1$]{$\hat{u} \in \mathbb{R}^d$}{Reynolds-averaged velocity}
\nomenclature[$n$]{$n_{\text{s}}$}{specific speed of the pump}
\nomenclature[$u2$]{$u_{\text{ref}} \in \mathbb{R}^d$}{reference velocity}
\nomenclature[$p$]{$p \in \mathbb{R}$}{pressure}
\nomenclature[$p1$]{$\hat{p} \in \mathbb{R}$}{Reynolds-averaged pressure}
\nomenclature[$p2$]{$p_{\text{ref}} \in \mathbb{R}$}{reference pressure}
\nomenclature[$d$]{$d$}{number of dimensions of the spatial domain}
\nomenclature[$x$]{$x$, $x_n$}{spatial location}
\nomenclature[$t$]{$t$, $t_m$}{time}
\nomenclature[$N$]{$N_{\text{grid}}$}{number of spatial locations on the grid}
\nomenclature[$N$]{$N$}{$= d N_\text{grid}$} 
\nomenclature[$N$]{$N_\text{EKF}$}{number of measurement locations}
\nomenclature[$M$]{$M$}{number of snapshots}
\nomenclature[$u3$]{$\bar{u} \in \mathbb{R}^d$}{time-averaged velocity}
\nomenclature[$u3$]{$\tilde{u} \in \mathbb{R}^d$}{time-variant velocity}
\nomenclature[$U$]{$\tilde{U} \in \mathbb{R}^{N\times M}$}{time-variant velocity snapshot matrix}
\nomenclature[$a$]{$a, a^{\text{opt}}, a^{\text{POD}}, a^{\text{EKF}}$}{time-variant velocity coefficients from the velocity ROM, optimized velocity ROM, POD, and EKF}
\nomenclature[$b$]{$b, b^{\text{opt}}, b^{\text{POD}}, b^{\text{EKF}}$}{time-variant pressure coefficients from the velocity ROM, optimized velocity ROM, POD, and EKF}
\nomenclature[$R$]{$R$}{number of reduced velocity POD basis vectors}
\nomenclature[$S$]{$S$}{surface area}
\nomenclature[$Q$]{$Q\in\mathbb{R}^{R\times R\times R}$}{coefficients of the quadratic term in the velocity ROM}
\nomenclature[$L$]{$L\in\mathbb{R}^{R\times R}$}{coefficients of the linear term in the velocity ROM}
\nomenclature[$C$]{$C\in\mathbb{R}^{R}$}{coefficients of the constant term in the velocity ROM}
\nomenclature[$p3$]{$\bar{p} \in \mathbb{R}$}{time-averaged pressure}
\nomenclature[$p3$]{$\tilde{p} \in \mathbb{R}$}{time-variant pressure}
\nomenclature[$P$]{$\tilde{P} \in \mathbb{R}^{N_{\text{grid}}\times M}$}{time-variant pressure snapshot matrix}
\nomenclature[$R$]{$R_{\text{p}}$}{number of reduced pressure POD basis vectors}
\nomenclature[$Q$]{$Q_{\text{p}}\in\mathbb{R}^{R_{\text{p}}\times R\times R}$}{coefficients of the quadratic term in the pressure ROM based on pressure and velocity modes}
\nomenclature[$C$]{$C_{\text{p}}\in\mathbb{R}^{R_{\text{p}}}$}{coefficients of the constant term in the pressure ROM based on pressure and velocity modes}
\nomenclature[$P$]{$P_Q\in\mathbb{R}^{N_{\text{grid}}\times R\times R}$}{coefficients in the pressure ROM based on velocity modes}
\nomenclature[$C$]{$\mathcal{C}\in\mathbb{R}^{dN_{\text{EKF}}\times R}$}{output matrix}
\nomenclature[$O$]{$\mathcal{O}\in\mathbb{R}^{dN_{\text{EKF}}R\times R}$}{observability matrix}
\nomenclature[$J$]{$J_f$}{Jacobian matrix}
\nomenclature[$y$]{$y\in\mathbb{R}^{dN_{\text{EKF}}}$}{velocity measurements}
\nomenclature[$K$]{$K \in \mathbb{R}^{R\times dN_{\text{EKF}}}$}{Kalman gain}
\nomenclature[$E$]{$\mathcal{E}_{\text{u},\text{TRU}}$}{velocity POD truncation error}
\nomenclature[$E$]{$\mathcal{E}_{\text{p},\text{TRU}}$}{pressure POD truncation error}
\nomenclature[$E$]{$\mathcal{E}_{\text{u},\text{ROM}}$}{velocity ROM approximation error}
\nomenclature[$E$]{$\mathcal{E}_{\text{p},\text{ROM}}$}{pressure ROM approximation error}
\nomenclature[$E$]{$\mathcal{E}_{\text{u},\text{REC}}$}{velocity reconstruction error}
\nomenclature[$E$]{$\mathcal{E}_{\text{p},\text{REC}}$}{pressure reconstruction error}

\nomenclature[A$11$]{$\Theta\in\mathbb{R}^{R\times R}$}{a priori and a posteriori estimate covariance}
\nomenclature[A$12$]{$\Theta_M\in\mathbb{R}^{dN_{\text{EKF}}\times dN_{\text{EKF}}}$}{measurement noise covariance}
\nomenclature[A$13$]{$\Theta_P\in\mathbb{R}^{R\times R}$}{process noise covariance}
\nomenclature[A$14$]{$\kappa$}{observability condition number}
\nomenclature[A$16$]{$\nu$}{kinematic viscosity}
\nomenclature[A$16$]{$\nu_t$}{kinematic eddy viscosity}
\nomenclature[A$17$]{$\Xi\in\mathbb{R}^{N_\text{EKF}}$, $\xi$}{measurement locations}
\nomenclature[A$18$]{$\rho$}{fluid density}
\nomenclature[A$19$]{$\Sigma\in\mathbb{R}^{M\times M}$, $\sigma$}{velocity singular values}
\nomenclature[A$21$]{$T\in\mathbb{R}^{M\times M}$, $\tau$}{pressure singular values}
\nomenclature[A$22$]{$\Phi\in\mathbb{R}^{N\times M}$, $\phi\in\mathbb{R}^d$}{spatial velocity POD basis vectors (velocity POD mode)}
\nomenclature[A$23$]{$\Psi\in\mathbb{R}^{N_{\text{grid}}\times M}$, $\psi\in\mathbb{R}$}{spatial pressure POD basis vectors (velocity POD mode)}
\nomenclature[A$24$]{$\Omega$}{spatial domain}

\nomenclature[B]{ROM}{reduced order model}
\nomenclature[B]{POD}{proper orthogonal decomposition}
\nomenclature[B]{EKF}{extended Kalman filter}
\nomenclature[B]{\textit{PISO}}{pressure-implicit with splitting of operators}
\nomenclature[B]{\textit{SIMPLE}}{semi-implicit method for pressure linked equations}
\nomenclature[B]{SST}{shear stress transport turbulence model}
\nomenclature[B]{TVD}{total variation diminishing}
\nomenclature[B]{URANS}{unsteady Reynolds-averaged Navier-Stokes}
\nomenclature[B]{3D}{three-dimensional}
\nomenclature[B]{2D}{two-dimensional}	
\nomenclature[B]{CFD}{computational fluid dynamics}
\nomenclature[B]{GGI}{general grid interface}

\begin{figure*}
    \begin{mdframed}
        \printnomenclature[2.7cm]
    \end{mdframed}
\end{figure*}